\documentclass[preprint,12pt]{elsarticle}
\usepackage{type1cm}
\usepackage{geometry}
\usepackage{eso-pic}
\usepackage[all]{xy}
\usepackage{multirow} 
\usepackage{url}
\usepackage{amsmath}
\usepackage{amssymb}
\usepackage{amsthm}

\newtheorem{theorem}{Theorem}[section]
\newtheorem{lemma}[theorem]{Lemma}

\theoremstyle{remark}
\newtheorem{remark}[theorem]{Remark}

\theoremstyle{definition}

\newcommand{\range}{\operatorname{Range}}
\newcommand{\R}{\mathbb{R}}

\newcommand{\tx}[1]{\textrm{#1}}

\renewcommand{\vec}[1]{\boldsymbol{#1}}
\newcommand{\trace}{\operatorname{trace}}
\renewcommand{\th}{^{\textrm{\footnotesize{th}}}}

\newcommand{\vectwo}[2]{ \left[\!\!
			     \begin{array}{c}
                             #1\\
			     #2 
                             \end{array} \!\! \right]}

\definecolor{darkgreen}{rgb}{0.09,0.44,0.04}
\definecolor{hellbrown}{rgb}{0.63,0.25,0}

\biboptions{square,comma}
\journal{XXX}
\begin{document}
 \begin{frontmatter}
\title{Computing Optimal Designs of multiresponse Experiments reduces to Second-Order Cone Programming}
\author{Guillaume Sagnol\fnref{fn:founding}}
\address{Inria Saclay \^Ile de France \&  Centre de Math\'ematiques Appliqu\'ees de l'\'Ecole Polytechnique (CMAP)\\ \url{guillaume.sagnol@inria.fr}}
\fntext[fn:founding]{The doctoral work of the author is supported by Orange Labs through the research contract CRE EB 257676 with INRIA.}

%\maketitle%ENLEVER A LA FIN 
%\todo{remettre frontmatter}

\begin{abstract}

Elfving's Theorem is a major result in the theory of optimal experimental design,
which gives a geometrical characterization of $\vec{c}-$optimality. In this
paper, we extend this theorem to the case of multiresponse experiments, and we show
that when the number of experiments is finite, the $\vec{c}-,A-,T-$ and $D-$optimal design of multiresponse experiments can be computed
by Second-Order Cone Programming (SOCP).
Moreover, the present SOCP approach can deal with design problems in which the variable is subject to
several linear constraints.

We give two proofs of
this generalization of Elfving's theorem. One is based on Lagrangian dualization techniques and relies on the
fact that the semidefinite programming (SDP)
formulation of the multiresponse $\vec{c}-$optimal design always has a solution which is a matrix of rank $1$. Therefore,
the complexity of this problem fades.

We also investigate a \emph{model robust} generalization of $\vec{c}-$optimality, for which an Elfving-type theorem was established
by Dette (1993).
We show with the same Lagrangian approach that these model robust designs
can be computed efficiently by minimizing a geometric mean under some norm constraints.
Moreover, we show that the optimality conditions of this geometric programming problem yield
an extension of Dette's theorem to the case of multiresponse experiments.

When the goal is to identify a small number of linear functions of the unknown parameter (typically for
$\vec{c}-$optimality),
we show by numerical examples that the present approach can be between 10 and 1000 times faster than the classic, 
state-of-the-art algorithms.
\end{abstract}


\end{frontmatter}

\section{Introduction}
%The theory of \textit{experimental design}
%has been developed since the 1920's, after some work
%of Gosset~\cite{Stud17} (known under th pseudonym ``Student'') and Fisher, 
%who introduced several useful concepts for a theoretical approach to the design of
%experiments in his book~\cite{Fish35}.
%We refer the reader to the article of Atkinson and Bailey~\cite{AB01} for a
%review on the early development of the theory of optimal experiments.
%
%
%The theory of \textit{experimental design} has been developed since the 1940's after the work of Sir A. Fisher~\cite{Fish47},
%and plays a central role in statistics. 

An important branch of statistics is the \emph{theory of optimal experimental designs}, which explains how to best select experiments in order to estimate
a vector of parameters $\vec{\theta}$. We refer the reader to the monographs of
Fedorov~\cite{Fed72} and Pukelsheim~\cite{Puk93} for a comprehensive review on the subject,
and to an article of Atkinson and Bailey~\cite{AB01} for
more details on the early development of this theory.

The importance of Elfving's Theorem (1952)~\cite{Elf52}, which was one of the first major improvements in the field of optimal design of experiments,
was illustrated in many works~\cite{Cher99,Dette93,DS93,DHS95,Stud71,Stud05}.
This result gives a geometrical characterization of the experimental design which minimizes the variance of the best estimator for
the linear combination of the parameters  $\vec{c}^T \vec{\theta}$. Namely, the optimal design can be found at the intersection of a vectorial straight-line and
the boundary of a
convex set referred as \emph{Elfving Set}.
When the number of available experiments is finite, Elfving set becomes a polyhedron, and so Elfving's geometrical characterization
allows one to compute $\vec{c}-$optimal designs by linear programming methods,
see Harman and Jur\`ik~\cite{HJ08}.\par\vspace{12pt}

An interesting case appears in the study of the design of experiments, when a single experiment 
is allowed to give simultaneously several observations of the parameters.
This setting is referred in the literature
as \emph{multiresponse experiments}, and occurs in many practical situations. For more details on the optimal design of
multiresponse experiments, the reader is referred to the book of Fedorov~\cite{Fed72}.\par\vspace{12pt}

In this paper, we extend Elfving's classic result (Theorem~\ref{theo:Elfving}) to the
optimal design of multiresponse experiments, see Theorem~\ref{theo:ElfvingExt}, and we show
that the latter problem reduces to a Second-Order Cone Programming problem
(SOCP), see Theorem~\ref{theo:SOCP}.

Moreover, we show in Section~\ref{sec:Ades} that this result generalizes the Elfving-type result
of Studden~\cite{Stud71}, which characterizes geometrically the $A-$optimal design for the estimation of a multidimensional linear
combination of the parameters. As a consequence, one may cast the problem of finding an $A-$optimal design
on a finite regression range (for single- or multi- response experiments) as an SOCP.

In contrast to the classic algorithms for the computation of optimal designs, the flexibility of mathematical programming
approaches makes it possible to add constraints in the problem \emph{without additional effort}. We will
see that the present SOCP approach can indeed handle optimal design problems subject to multiple resource constraints, see Section~\ref{sec:multCons}.

The fact that the $\vec{c}-$optimal design problem (on a finite regression space) has a semidefinite programming (SDP) formulation can be traced back to 1980,
as a particular case of Theorem~4
in Pukelsheim~\cite{Puk80}. Similarly, the classic
A- and D- optimal design problems for multiresponse experiments can be formulated as SDP~\cite{VBW98}. Second Order Cone Programming (SOCP)
is a class of convex programs which is somehow harder than linear programming (LP),
but which can be solved by interior point codes like SeDuMi~\cite{sedumi} in a much shorter time that semidefinite programs (SDP) of the same size.
Moreover the SOCP method takes advantage
of the sparsity of the matrices arising in the problem formulation. 
Hence, this work shows that computing $A$- and $\vec{c}-$optimal designs actually belongs to an easier class of problems, 
and makes it possible to solve instances that were previously intractable.

While our proof of Theorem~\ref{theo:ElfvingExt} is an extension of Elfving's original proof, it leaves unexplained why the SDP formulation actually reduces
to an SOCP. We provide in Section~\ref{sec:lagrange}
another proof of the present Elfving-type result based on
Lagrangian relaxation, which explains why the complexity of this problem fades.
The proof relies on Theorem~\ref{theo:SDPSol1}, which shows
that a certain class of semidefinite programs (packing programs
with a rank-one objective function) have a rank-one solution.
Theorem~\ref{theo:SDPSol1} appears to be a result of an independent interest,
in relation with the study of semidefinite relaxations of combinatorial
optimization problems, and is therefore the subject of the companion paper~\cite{Sagnol09rank}.\par\vspace{12pt}

We next show (Theorem~\ref{theo:Topt}) that $T-$optimality also admits a
second order cone programming representation,
which gives another argument for saying that second order cone programming is a natural tool for
handling experimental optimal design problems. If the experimenter wishes to estimate the
full vector of parameters $\vec{\theta}$, the $T-$optimal design problem is trivial.
However, if he is interested in a linear subsystem $K^T\vec{\theta}$,
the $T-$optimal design problem is complicated and can be handled by SOCP.

\sloppypar{In Section~\ref{sec:Sopt}, we consider a generalization of $\vec{c}-$optimality which was originally proposed by L\"auter~\cite{Laut74} 
in order to deal with the uncertainty on the
model. In this approach (called $S-$optimality by L\"auter), 
the objective criterion
takes into account the variance of several estimators, balanced in a log term with coefficients which indicate the belief of the
experimenter in each model. Dette~\cite{Dette93} characterized by an Elfving-type result the $S-$optimal designs.
We  show in Theorem~\ref{theo:SoptMulti}, which is the main result of this section, that when the number of available experiments is finite,
$S-$optimal designs of multiresponse experiments can be computed efficiently by minimizing a geometric mean under some norm constraints.
Moreover, we show in Theorem~\ref{theo:DetteMulti} that the optimality conditions of this geometric
program yield
an extension of Dette's theorem to the case of multiresponse experiments. 
As a consequence of Theorem~\ref{theo:SoptMulti}, we obtain a SOCP for $D-$optimality (cf.\ Remark~\ref{rem:Dopt}).
The results of this section are proved in appendix.}\par\vspace{12pt}

This work grew out from an application to networks~\cite{BGSagnol08Rio},
in which the traffic between any two pairs of nodes must
be inferred from a set of measurements. This leads to a large
scale optimal experimental design problem which can not be handled by standard SDP solvers.
In a companion work relying on the present reduction to an SOCP~\cite{SagnolGB10ITC}, 
we illustrate our method by solving within seconds optimal design problems that could not be handled by SDP.
In addition, the constraints of the SOCP formulation involves the \textit{observation matrices} $A_i$ of the experiments, which happen to be very sparse in practice.
This is in contrast with the \textit{information matrices} involved in the SDP formulation ($M_i=A_i^T A_i$), which are not sparse in general. The present
SOCP formulation takes advantage of the intrinsic sparsity of the data, and the instances can be solved very efficiently.

We present in Section~\ref{sec:numexp} some numerical experiments for problems of the kind that arise in network monitoring, as well as
classic polynomial regression problems.
Our experiments show that the SOCP approach is well suited when the number of linear functions to estimate is \emph{small}.
For the case of $\vec{c}-$optimality (only one linear function to estimate), we will see that solving the second order cone program
is usually 10 times faster than the classic exchange or multiplicative algorithms, and up to 2000 times faster than the SDP approach
for problems with multiple constraints.\par\vspace{12pt}

Some results of this paper, including Theorem~\ref{theo:ElfvingExt}, were presented at the conference~\cite{SagnolBG09Pisa}, and the technical result justifying the reduction to a SOCP was posted on
arXiv~\cite{Sagnol09rank}. Shortly before the time of submission, Dette and Holland-Letz published
an article in \emph{Annals of Statistics}, in which Theorem~\ref{theo:ElfvingExt} was established independently (Theorem $3.3$ in~\cite{DH09}). Dette and Holland-Letz considered a heteroscedastic model (i.e.\ an experimental
model where both the mean and the variance of the
observations depend on the parameter of interest), which led them to study the case in which the observation matrices are of rank
$k\geq 2$, just as in the model of \emph{multiresponse experiments}. They used their geometrical characterization of the
$\vec{c}-$optimal design for heteroscedastic models in an application to toxicokinetics and pharmacokinetics. It should also be
mentioned that the proof of Dette and Holland-Letz relies on an equivalence theorem (Theorem $3.1$ in~\cite{DH09}), while ours is closer to Elfving's original approach, as done previously by Studden~\cite{Stud05} for other results in optimal design
of experiments. The main result of this article (reduction to a SOCP, Theorem~\ref{theo:SOCP}), provides a new insight on the relations between these two approaches : they are actually
dual from each other (in the Lagrangian sense). Indeed, the approach of Dette and Holland-Letz corresponds to
the primal SOCP~\eqref{P-SOCP}, while our geometrical characterization corresponds to the dual SOCP~\eqref{D-SOCP},
 and strong duality holds between these two optimization problems.

%Note that Theorem~\ref{ElfvingExt} was presented at the conference~\cite{SagnolBG09Pisa}
%in April 2009. Shortly before the time of submission, Dette and Holland-Letz published an article where this result was established independently (Theorem $3.3$ in~\cite{DH09}).
%The proof presented in this article is different from Dette and Holland-Letz's, though, and is inspired from the original idea of Elfving\cite{Elf52}, as reformulated by Studden~\cite{Stud05} with more recent notations.

\section{Preliminary results}

Before stating the main results of this paper, we introduce the necessary
background and recall Elfving's classic result. Throughout this paper, we denote vectors by boldface letters and matrices by capital letters.
We make use of the standard notation $[n]=\{1,\ldots,n\}$. The components of a vector $\vec{v}\in \R^n$
are denoted by $v_1,\ldots,v_n$. The $L^2-$norm of $\vec{v}$ is denoted by $\Vert \vec{v} \Vert$, and the
Frobenius norm of a matrix $A$ by $\Vert A \Vert_F =\sqrt{\trace\ (A A^T)}$. The notation $\vec{v}\geq\vec{0}$
means that every component of $\vec{v}$ is nonnegative.

The most common model in optimal experimental design assumes
that each experiment provides a measurement which is a linear combination of
the parameters up to the accuracy of the measurement. Let $\mathcal{X}$ denote the set of available experiments.
Every experiment $\vec{x}\in \mathcal{X}$ provides a measurement
\begin{equation}
y(\vec{x})=\vec{a_x}^T \vec{\theta}+\epsilon, \label{obsequations}
\end{equation}
where $\vec{\theta}$ is the $m-$dimensional vector of unknown parameters, and $\vec{a_x}$ is the ($m \times 1$)-regression
vector. Uncorrelated experiments are performed at different locations $\vec{x_1},...,\vec{x_s}$
from the compact set $\mathcal{X}$ (possibly infinite), and the objective is to determine both the optimal 
choice of the $\vec{x_i}$, and the number of experiments $n_i$ to be conducted at $\vec{x_i}$ ; we call such a subset of experiments a \emph{design}.

Rather than deciding the exact number of times that each experiment should be conducted,
it has been proposed to work with \textit{approximate designs} instead,
which is simply done by releasing the integrity constraints on the $n_i$.
In this setting, a mass indicates the proportion from the total number of experiments to be
conducted for each available experiment. For example, if the weight for the $i^{\tx{th}}$ 
experiment is $w_i$, and that $N$ experiments are allowed, $Nw_i$ are chosen at $\vec{x_i}$.
This definition suggests that the vector of weights $\vec{w}$ is such that each quantity $Nw_i$ is integer. However, the
continuous relaxation where every vector $\vec{w}$ summing to $1$ is allowed is of theoretical interest.
The approximate design where the \emph{percentage of experimental effort} 
at $\vec{x_i}$ is $w_i$ is written as $$\xi=\left(\begin{array}{ccc}
\vec{x_1} & \cdots  & \vec{x_s}\\
w_1 & \cdots & w_s \end{array} \right),$$
or $\xi=\{\vec{x_i},w_i\}$ for short. In this paper, we consider only approximate designs.\par\vspace{12pt}

The celebrated result of Elfving~\cite{Elf52} gives a geometrical characterization of the design (known as $\vec{c}-$optimal) which
minimizes the variance of an unbiased estimator for a single linear combination $\vec{c}^T \vec{\theta}$ of the parameters. In this paper, 
we are going to extend Elfving's Theorem to the multiresponse case. To this end, we consider the linear regression model
$$\vec{y}(\vec{x})=A(\vec{x}) \vec{\theta} + \vec{\epsilon}(\vec{x}),\quad \vec{x} \in \mathcal{X},$$ where 
$\vec{\theta}$ is an unknown $m$-dimensional parameter, $\vec{y}(\vec{x})$ is the $l$-dimensional measurement vector
for the experiment at $\vec{x}$,  $A(\vec{x})$ is a $l \times m$ observation matrix, and $\epsilon(\vec{x})$
is the error on the measurement. The matrix $A(\vec{x})$ may eventually become rank deficient for certain values of $\vec{x}$.
This allows one to handle the case in which the experiment at $\vec{x}$ gives only $l_{\vec{x}}<l$ measurements. In this case, we set 
$l-l_{\vec{x}}$ rows of the matrix $A(\vec{x})$ to zero.

The observations are uncorrelated and have a constant variance, which will be assumed to be one for the ease of the presentation, i.e.\
$$\forall \vec{x_1},\vec{x_2} \in \mathcal{X},\ \vec{x_1} \neq \vec{x_2},\ \mathbb{E}(\vec{\epsilon}(\vec{x_1}))=\vec{0},\quad \mathbb{E}(\vec{\epsilon}(\vec{x_1}) \vec{\epsilon}(\vec{x_1})^T)=\vec{I},\quad \mathbb{E}(\vec{\epsilon}(\vec{x_1}) \vec{\epsilon}(\vec{x_2})^T)=0.$$
In fact, one can always reduce to this case when the variance of the experiments is known, after
a left scaling of the observation equations.

When $n_i=Nw_i$ experiments are conducted at $\vec{x_i}$, we denote by $\overline{\vec{y}}(\vec{x_i})$ the average of these observations:
%$\overline{Y(x_i)}=\frac{1}{n_i} (Y(x_i)_1+...+Y(x_i)_{n_i} )$, so that 
we have $\mathbb{E}(\overline{\vec{y}}(\vec{x_i}))=A(\vec{x_i}) \vec{\theta}$, and $\operatorname{Var}(\overline{\vec{y}}(\vec{x_i}))=\frac{1}{n_i} \vec{I}$.
For the design $\xi=\{\vec{x_i},w_i\}$, we denote by $\vec{y}(\xi)$ the aggregate vector of observations $[\overline{\vec{y}}(\vec{x_1})^T,...,\overline{\vec{y}}(\vec{x_s})^T]^T$, and by $A(\xi)$ the aggregate observation matrix $[A(\vec{x_1})^T,...,A(\vec{x_s})^T]^T$,
so that $\mathbb{E}(\vec{y}(\xi))=A(\xi) \vec{\theta}$, and $\operatorname{Var}(\vec{y}(\xi))=\frac{1}{N} \Delta(\vec{w})$, where
\begin{equation}
 \Delta(\vec{w})=\left( \begin{array}{ccc}
                 w_1^{-1} \vec{I} & &\\
		& \ddots &\\
		& & w_s^{-1} \vec{I}
                \end{array} \right), \label{Deltaw}
\end{equation}
with $(l \times l)-$identity blocks on the diagonal. If $w_i=0$ for some $i\in[s]$, 
we simply remove the measurement point $\vec{x_i}$ from $\xi$. For ease of presentation, we get rid of the multiplication factor $1/N$, since it
does not affect the results on optimal designs.\par\vspace{12pt}

Assume now that an experimenter wishes to estimate the scalar quantity $\zeta=\vec{c}^T \vec{\theta}$, that is to say that he wants to estimate a linear
combination of the parameters. It can easily be seen that a linear estimator $\hat{\zeta}=\vec{h}^T \vec{y}(\xi)$ is
unbiased if and only if $A(\xi)^T \vec{h}=\vec{c}$. Thus, linear unbiased estimators for
$\vec{c}^T \vec{\theta}$ exist as long as
$\vec{c}\in \operatorname{Range}(A(\xi)^T)$. We will say that the quantity $\zeta=\vec{c}^T\vec{\theta}$ is
\emph{estimable} when there is a design $\xi$ such that $\vec{c}\in\range(A(\xi)^T)$.
Notice that a sufficient condition 
which ensures that $\vec{c}^T\vec{\theta}$ is estimable for all $\vec{c}\in\R^m$ is that
the matrices $\big(A(\vec{x})\big)_{\vec{x}\in\mathcal{X}}$ contain $m$ linearly independent vectors
among their rows. For an estimable quantity $\vec{c}^T\vec{\theta}$,
we define the feasibility cone $\Xi(\vec{c})$ as 
the set of designs $\xi$ such that $(A(\xi)^T)$ span the vector $\vec{c}$, and a design $\xi$ will be said
\emph{feasible} if it lies in the feasibility cone.

Now, let us assume that the quantity $\vec{c}^T\vec{\theta}$ is estimable. Our protagonist certainly wants to choose an unbiased estimator of $\zeta$ with minimal variance;
the variance of $\hat{\zeta}$ is equal to $\vec{h}^T \operatorname{Var}(\vec{y}(\xi)) \vec{h}=\vec{h}^T \Delta(\vec{w}) \vec{h}$, and 
we know from Gauss-Markov theorem that it is minimized under the unbiasedness
constraint $A(\xi)^T \vec{h}=\vec{c}$ for $\vec{h}=\Delta(\vec{w})^{-1}A(\xi)(A(\xi)^T \Delta(\vec{w})^{-1} A(\xi))^\dagger \vec{c}$,
where $M^\dagger$ denotes the Moore-Penrose pseudo inverse of $M$, and the variance
of this best linear unbiased estimator is :
$$\textrm{Var}(\hat{\zeta})=\vec{c}^T (A(\xi)^T \Delta(\vec{w})^{-1} A(\xi))^\dagger \vec{c}=\vec{c}^T M(\xi)^- \vec{c},$$
\sloppypar{\noindent where $M(\xi)^-$ is a \textit{generalized inverse} of $M(\xi)=A(\xi)^T \Delta(\vec{w})^{-1} A(\xi)$, i.e.\ any matrix $G$ verifying
$M(\xi)GM(\xi)=M(\xi)$. Notice that this expression does not depend on the choice of the generalized inverse $G$, since
there exists a vector $\vec{u}$ such that $M(\xi)\vec{u}=\vec{c}$: }
$$\forall G \in M(\xi)^-,\quad \vec{c}^T G\vec{c}=\vec{u}^T M(\xi)GM(\xi)\vec{u}=\vec{u}^T M(\xi)\vec{u}.$$ 
The $(m \times m)-$matrix $M(\xi)$ is traditionally called the information matrix of the design, and it can be written as a weighted sum
of the observation blocks :
\begin{equation}
 M(\xi)=\sum_{i=1}^s w_i A(\vec{x_i})^T A(\vec{x_i}). \label{infoMat}
\end{equation}
In a more general setting, $\vec{w}$ is replaced by a probability measure $\mu$ on $\mathcal{X}$ :
$$M(\xi)=\int_\mathcal{X} A(\vec{x})^T A(\vec{x}) d\mu(x).$$ However, this continuous form of the information matrix
is still a symmetric matrix from the closed
convex hull of $\{A(\vec{x})^TA(\vec{x}),\vec{x}\in \mathcal{X}\}$. When $\mathcal{X}$ is compact, and $\vec{x} \mapsto A(\vec{x})$ is continuous, the set
of all information matrices $\{A(\vec{x})^TA(\vec{x}),\vec{x}\in \mathcal{X}\}$ is closed, and we know from Caratheodory's theorem that $M(\xi)$
can be written as barycenter of $m(m+1)/2+1$ information matrices (see Fedorov~\cite{Fed72}).
Therefore, the optimal design can always be expressed with a discrete measure
$\mu=w_1 \delta(\vec{x}-\vec{x_1})+...+w_s \delta(\vec{x}-\vec{x_s})$,
where $\delta$ is the Dirac measure and $s\leq m(m+1)/2+1$. We will consider
only such discrete designs in this work. \par\vspace{12pt}

The $\vec{c}-$optimal problem is to find the feasible design $\xi=\{\vec{x_i},w_i\}$ minimizing the variance
of the aforementioned estimator:
\begin{align} \min_{\xi\in\Xi(\vec{c})}\ \ & \vec{c}^T M(\xi)^- \vec{c} \label{mincMc}\\
\operatorname{s.t.}\ & M(\xi)=\sum_{i=1}^s w_i A(\vec{x_i})^T A(\vec{x_i}) \nonumber\\
&  \sum_{i=1}^s w_i=1;\quad \forall\ i \in [s], w_i\geq 0, \vec{x_i}\in \mathcal{X}. \nonumber
\end{align}
We point out that Problem~\eqref{mincMc} is not limited to the case of multiresponse experiments;
there are other
situations where the information matrix takes the form of Equation~\eqref{infoMat} (with $l>1$), for example
in the case of models with parametrized variance
or in models with correlated observations (see e.g.~\cite{MP95,Paz04}).

In the classic model (single response experiments), $y(\vec{x})$ is a scalar observation,
which means that $l=1$ and $A(\vec{x})=\vec{a_x}^T$ is a row vector. In this case, Elfving's Theorem gives a geometrical
characterization of the $\vec{c}-$optimal design. We first define the \textit{Elfving set} as the convex hull of
the vectors $\pm \vec{a_x}$:
$$\mathcal{E}=\operatorname{convex-hull} \big\{ \pm \vec{a_x},\  \vec{x}\in \mathcal{X}\big\},$$
and we denote its boundary by $\partial \mathcal{E}$.

\begin{theorem}[Elfving~{\cite{Elf52}}]
\label{theo:Elfving}
A design $\xi=\{\vec{x_i},w_i\}$ is $\vec{c}-$optimal if and only if there exists scalars $\epsilon_i=\pm 1$ and a positive real $t$ such that
$$t \vec{c} = \sum_{i=1}^s w_i \epsilon_i \vec{a_{x_i}} \in \partial \mathcal{E}.$$
Moreover, $t^{-2}=\vec{c}^T M(\xi)^- \vec{c}$ is the minimal variance.
\end{theorem}

%The generalization to multiresponse experiments that we give
%in Section~\todo{ref} has a proof relying on original ideas of Elfving, and so we will only proove this generalization (Theorem~\todo{ref}).
Elfving's theorem shows that the $\vec{c}-$optimal design is characterized by the intersection between the vectorial straight line directed by
$\vec{c}$ and the boundary of the Elfving set $\mathcal{E}$. We also point out that when the vector $\vec{c}$ is not spanned by the
regression vectors $(\vec{a_{\vec{x}}})_{\vec{x} \in \mathcal{X}}$, in other words when $\vec{c}^T\vec{\theta}$ is not estimable
(i.e.\ $\Xi(\vec{c})=\emptyset$), then the only scalar $t$ such that $t\vec{c}$ lies in $\mathcal{E}$ is $0$, 
and so a  $\vec{c}-$optimal design does not exist.

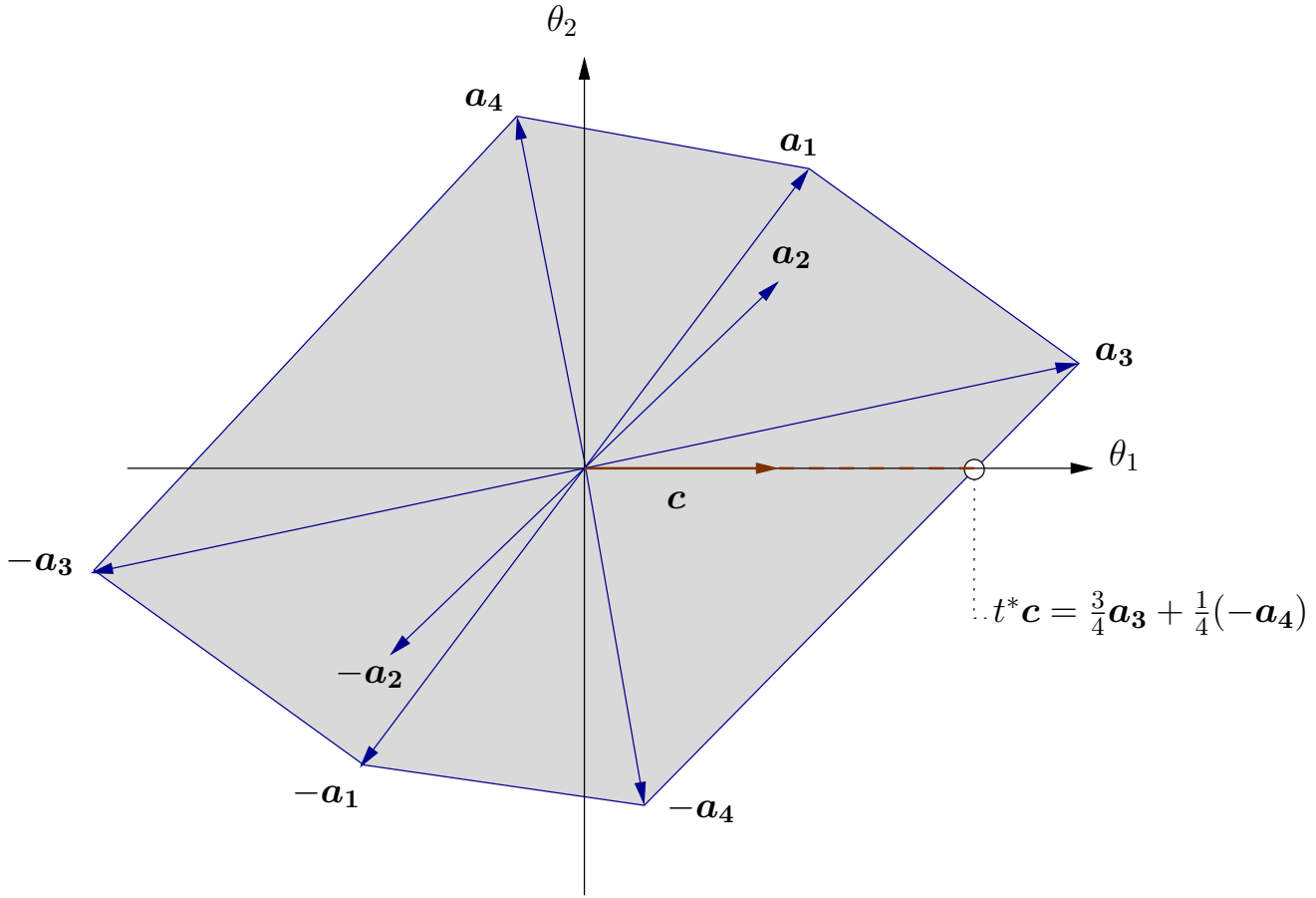
\begin{figure}[t!]
%---------------------------------------------------
% Xfig input

\begin{flushleft}
\begin{picture}(0,0)%
\includegraphics[scale=1]{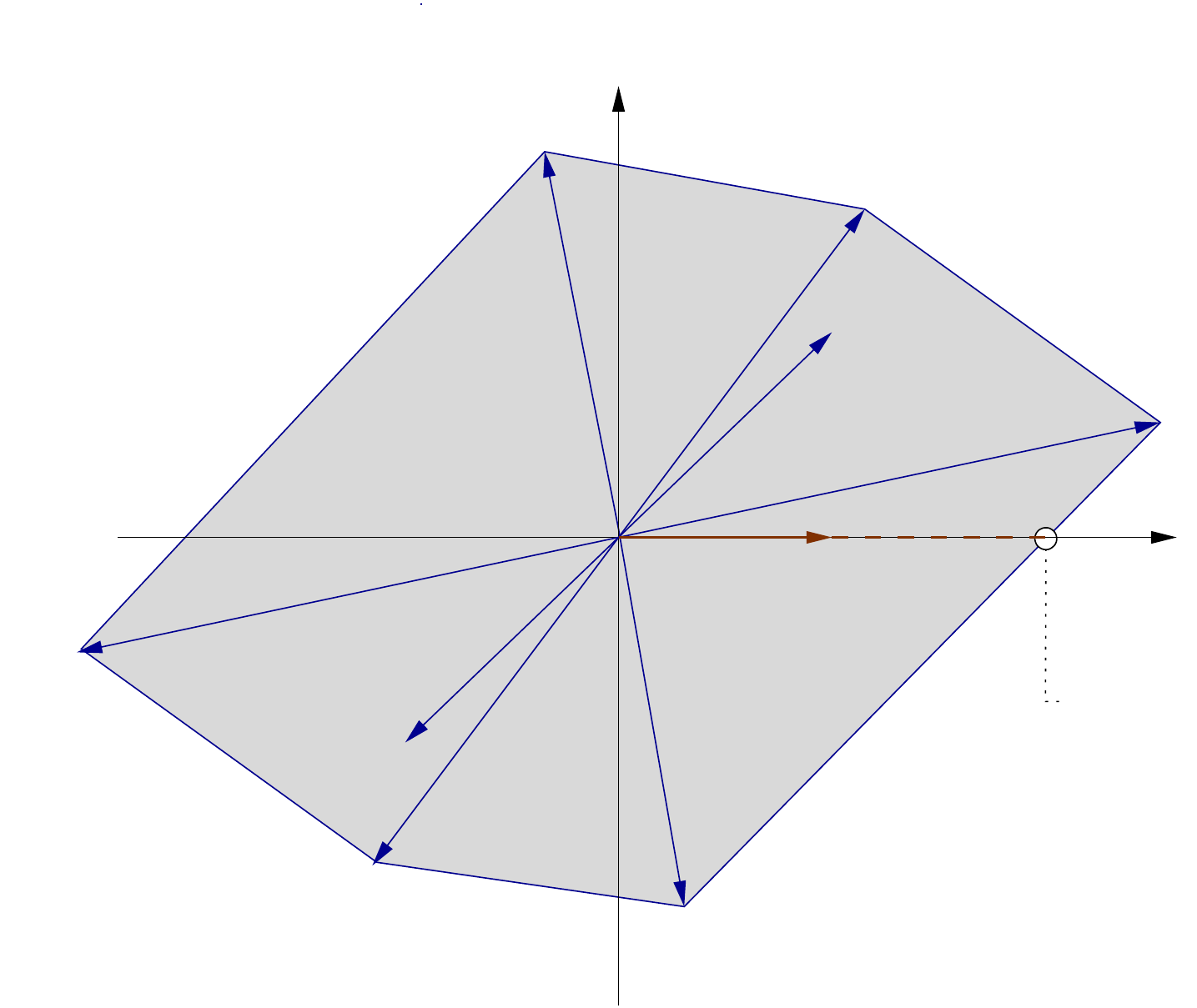}%
\end{picture}%
\setlength{\unitlength}{4144sp}%
\begingroup\makeatletter\ifx\SetFigFont\undefined%
\gdef\SetFigFont#1#2#3#4#5{%
  \reset@font\fontsize{#1}{#2pt}%
  \fontfamily{#3}\fontseries{#4}\fontshape{#5}%
  \selectfont}%
\fi\endgroup%
\begin{picture}(6555,4990)(166,-4728)
\put(4726,-241){\makebox(0,0)[lb]{\smash{{\SetFigFont{14}{16.8}{\familydefault}{\mddefault}{\updefault}{\color[rgb]{0,0,0}$\vec{a_1}$}%
}}}}
\put(4681,-916){\makebox(0,0)[lb]{\smash{{\SetFigFont{14}{16.8}{\familydefault}{\mddefault}{\updefault}{\color[rgb]{0,0,0}$\vec{a_2}$}%
}}}}
\put(1801,-4156){\makebox(0,0)[lb]{\smash{{\SetFigFont{14}{16.8}{\familydefault}{\mddefault}{\updefault}{\color[rgb]{0,0,0}$-\vec{a_1}$}%
}}}}
\put(4051,-2391){\makebox(0,0)[lb]{\smash{{\SetFigFont{14}{16.8}{\familydefault}{\mddefault}{\updefault}{\color[rgb]{0,0,0}$\vec{c}$}%
}}}}
\put(2061,-3436){\makebox(0,0)[lb]{\smash{{\SetFigFont{14}{16.8}{\familydefault}{\mddefault}{\updefault}{\color[rgb]{0,0,0}$-\vec{a_2}$}%
}}}}
\put(2836, 29){\makebox(0,0)[lb]{\smash{{\SetFigFont{14}{16.8}{\familydefault}{\mddefault}{\updefault}{\color[rgb]{0,0,0}$\vec{a_4}$}%
}}}}
\put(81,-2761){\makebox(0,0)[lb]{\smash{{\SetFigFont{14}{16.8}{\familydefault}{\mddefault}{\updefault}{\color[rgb]{0,0,0}$-\vec{a_3}$}%
}}}}
\put(3331,479){\makebox(0,0)[lb]{\smash{{\SetFigFont{14}{16.8}{\familydefault}{\mddefault}{\updefault}{\color[rgb]{0,0,0}$\theta_2$}%
}}}}
\put(6011,-3076){\makebox(0,0)[lb]{\smash{{\SetFigFont{14}{16.8}{\familydefault}{\mddefault}{\updefault}{\color[rgb]{0,0,0}$t^*\vec{c}=\frac{3}{4}\vec{a_3}+\frac{1}{4}(-\vec{a_4})$}%
}}}}
\put(6621,-1501){\makebox(0,0)[lb]{\smash{{\SetFigFont{14}{16.8}{\familydefault}{\mddefault}{\updefault}{\color[rgb]{0,0,0}$\vec{a_3}$}%
}}}}
\put(6706,-2131){\makebox(0,0)[lb]{\smash{{\SetFigFont{14}{16.8}{\familydefault}{\mddefault}{\updefault}{\color[rgb]{0,0,0}$\theta_1$}%
}}}}
\put(4051,-4246){\makebox(0,0)[lb]{\smash{{\SetFigFont{14}{16.8}{\familydefault}{\mddefault}{\updefault}{\color[rgb]{0,0,0}$-\vec{a_4}$}%
}}}}
\end{picture}%
\end{flushleft}
%\begin{center}
%\includegraphics[width=\textwidth]{ElfvingPDF.pdf}
%\end{center}
\caption{Geometrical representation of Elfving's theorem in dimension two. The gray area represents the Elfving set, which is
a polyhedron because $\mathcal{X}$ is finite (here, $\mathcal{X}=\{1,2,3,4\}$).
The intersection $t^*\vec{c}$ determines the
weights of the $\vec{c}-$optimal design: $\vec{w^*}=[0,0,\frac{3}{4},\frac{1}{4}]^T$.}
\label{fig:Elfving}
\end{figure}

We show on Figure~\ref{fig:Elfving} a representation of Elfving's theorem in dimension $2$. Here, $\mathcal{X}=\{1,2,3,4\}$ is finite, such that
the Elfving set is a polyhedron. The vector $\vec{c}$ is along the $\theta_1-$axis,
which means that the experimenter wants to estimate $\zeta=\theta_1$. The intersection between this axis and the Elfving set
indicates the optimal weights of the $\vec{c}-$optimal design: $w_3=\frac{3}{4}$ and $w_4=\frac{1}{4}$. Note that since
$\vec{a_2}$ is in the interior of the Elfving set, the experiment $2$ is never selected, whatever is the vector $\vec{c}$.
This example also shows that the optimal design $\vec{w^*}$ can be computed by linear programming (LP) when $\mathcal{X}$ is
finite (intersection of a straight line and a polyhedron).  This feature was noticed by
Harman and Jur\'ik~\cite{HJ08}, who formulated the $\vec{c}-$optimality LP:
\begin{align}
 \max_{t,\vec{w}}\ & \quad t\\
s.t.\quad &  t\vec{c} = \sum_i \epsilon_i \vec{a_i} \nonumber \\
 & -w_i \leq \epsilon_i \leq w_i, \quad \forall i \in [s] \nonumber\\
 & \sum_i w_i=1, \vec{w}\geq\vec{0}. \nonumber
\end{align}

\section{Extension to the case of multiresponse experiments}
\label{sec:multiresponse}
In this paper, we extend Elfving's result to the case of multidimensional observations, by
defining an analog of the Elfving set for the multiresponse case:
$$\mathcal{\overline{E}}=\operatorname{convex-hull} \big\{ A(\vec{x})^T \vec{\epsilon},\ \vec{\epsilon} \in \R^l,\
 \Vert \vec{\epsilon} \Vert \leq 1,\ \vec{x}\in \mathcal{X} \big\}.$$
Following our proof, we further show in Theorem~\ref{theo:SOCP} that
the $\vec{c}-$optimal design of multiresponse experiments can be formulated
as a Second Order Cone Program.

\begin{theorem}[Extension of Elfving's theorem for multiresponse experiments]
\label{theo:ElfvingExt}
A design $\xi=\{\vec{x_i},w_i\}$ is $\vec{c-}$optimal if and only if there exists a positive scalar $t$ and vectors $\vec{\epsilon_i}$ in the
unit ball of $\R^{l}$ (i.e.\ $\Vert \vec{\epsilon_i} \Vert \leq1$),
such that $$t \vec{c} = \sum_i w_i A(\vec{x_i})^T \vec{\epsilon_i} \in \partial \mathcal{\overline{E}}.$$
Moreover, $t^{-2}=\vec{c}^T M(\xi)^- \vec{c}$ is the minimal variance.
\end{theorem}

\begin{proof}
We consider an unbiased linear estimator for $\zeta=\vec{c}^T \vec{\theta}$ :
$$\hat{\zeta}=\vec{h}^T \vec{y}(\xi), \textrm{ with } \vec{h}=[\vec{h_1}^T,...,\vec{h_s}^T]^T \in \R^{sl},
 \quad \vec{h_i} \in \R^{l}.$$
The unbiasedness property forces the following equality to hold : $$A(\xi)^T \vec{h}=\sum_{i=1}^s A(\vec{x_i})^T \vec{h_i}=\vec{c}.$$
Now, the Cauchy-Schwarz inequality gives the following lower bound for the variance of~$\hat{\zeta}$ :
\begin{equation}
\operatorname{Var}(\hat{\zeta})=\vec{h}^T \Delta(\vec{w}) \vec{h}=\sum_{k=1}^s \frac{\Vert \vec{h_k} \Vert^2}{w_k}
\geq \big( \sum_{k=1}^s \Vert \vec{h_k} \Vert \big)^2, \label{VarLB}
\end{equation}
where and $\Delta(\vec{w})$ was defined in Equation~\eqref{Deltaw}. We recall that
we assume $\vec{w}>\vec{0}$ without loss of generality, since an experiment with a zero weight can be removed from the design $\xi$.

We show that $\frac{\vec{c}}{\sum_k \Vert \vec{h_k} \Vert} \in \mathcal{\overline{E}}$, by writing:
\begin{equation*}
\frac{\vec{c}}{\sum_k \Vert \vec{h_k} \Vert}=\frac{A(\xi)^T \vec{h}}{\sum_k \Vert \vec{h_k} \Vert}
=\sum_i A(\vec{x_i})^T \frac{\vec{h_i}}{\sum_k \Vert \vec{h_k} \Vert}=
\sum_{\{i:\Vert \vec{h_i} \Vert >0\}} \mu_i A(\vec{x_i})^T \epsilon_i, 
\end{equation*}
where $\mu_i=\frac{\Vert \vec{h_i} \Vert}{\sum_k \Vert \vec{h_k} \Vert}$ and $\epsilon_i=\frac{\vec{h_i}}{\Vert \vec{h_i} \Vert}$,
so that $\Vert\vec{\epsilon_i}\Vert=1$, $\mu_i \geq 0$  and $\sum_i \mu_i=1$.

Let $t$ be a positive scalar such that $t\vec{c}\in \partial \mathcal{\overline{E}}$. The fact that $\frac{\vec{c}}{\sum_k \Vert \vec{h_k} \Vert}
\in \mathcal{\overline{E}}$ implies 
\begin{equation}
\frac{1}{\sum_k \Vert \vec{h_k} \Vert} \leq t \Longrightarrow \big( \sum_{k=1}^s \Vert \vec{h_k} \Vert \big)^2 \geq t^{-2}. \label{ineqboarder}
\end{equation}
Combining~\eqref{VarLB} and~\eqref{ineqboarder} leads to the lower bound $t^{-2}$ for the variance of any linear unbiased estimator of $\zeta$.\par\vspace{12pt}

We will show that this lower bound is attained if and only if the design $\xi$ satisfies the condition of the theorem. To do this, notice that
for a design $\xi$ and an estimator $\vec{h}^T \vec{y}(\xi)$ to be optimal, it is 
necessary and sufficient that the inequalities~\eqref{VarLB} and~\eqref{ineqboarder} are equalities.
The Cauchy-Schwarz inequality~\eqref{VarLB}
is an equality if and only if $\vec{w}$  is proportional to the vector $[\Vert \vec{h_1} \Vert, ..., \Vert \vec{h_s}\Vert]^T,$ i.e.\
$$w_i=\frac{\Vert \vec{h_i} \Vert}{\sum_k \Vert \vec{h_k} \Vert}.$$
The second inequality~\eqref{ineqboarder} is an equality whenever $\frac{\vec{c}}{\sum_k \Vert \vec{h_k} \Vert}
\in \partial \mathcal{\overline{E}}$, i.e.\ $\frac{1}{\sum_k \Vert \vec{h_k} \Vert}=t,$
where $t$ is the largest real such that $t\vec{c} \in  \mathcal{\overline{E}}$. We can write
$$\partial \mathcal{\overline{E}} \ni t\vec{c}=t \sum_i A(\vec{x_i})^T \vec{h_i}=\sum_{\{i: \Vert \vec{h_i} \Vert >0\}}
\mu_i A(\vec{x_i})^T \vec{\epsilon_i},$$
with $\mu_i=t \Vert \vec{h_i} \Vert$ and $\vec{\epsilon_i}=\frac{\vec{h_i}}{\Vert \vec{h_i} \Vert}$. We have $\Vert\vec{\epsilon_i}\Vert=1$,
and the equality conditions are satisfied if and only if $\mu_i=w_i$. 
\end{proof}

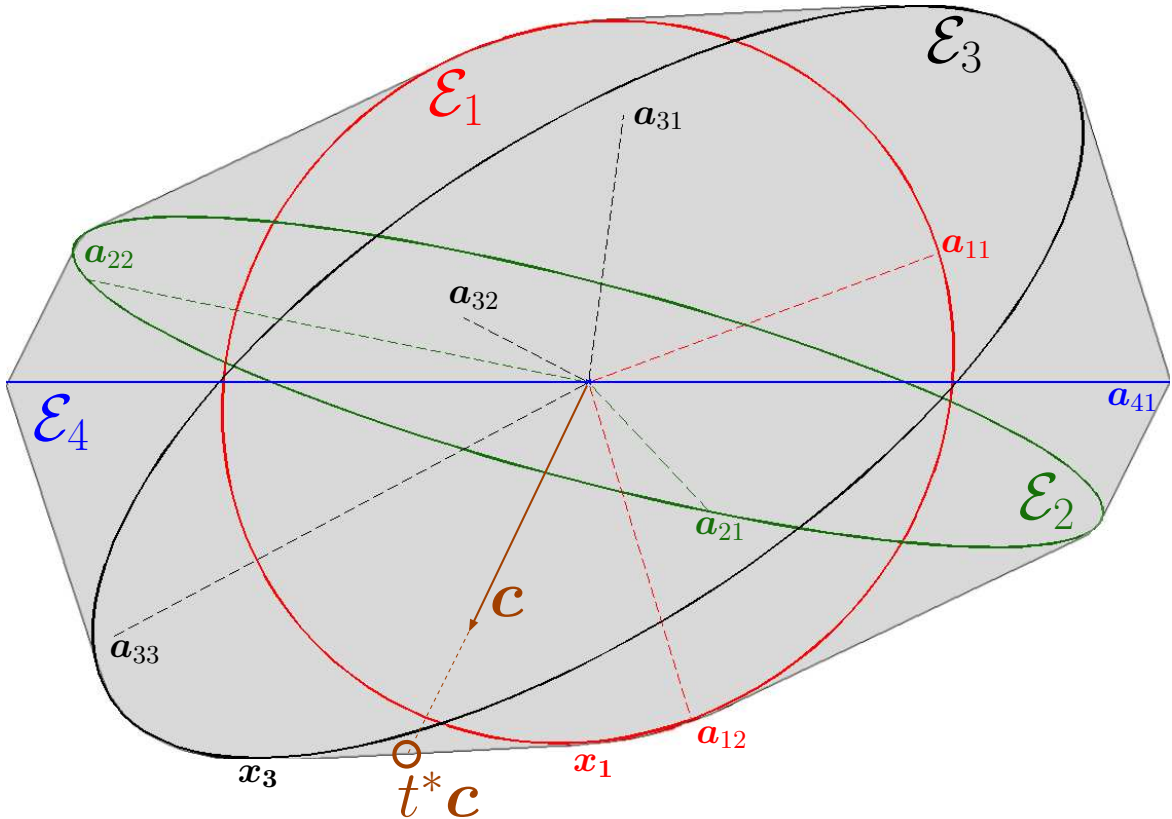
\begin{figure}[t!]
\begin{center}

% -----------------------------------------------
% Xfig Input

\begin{picture}(0,0)%
\includegraphics[scale=0.43]{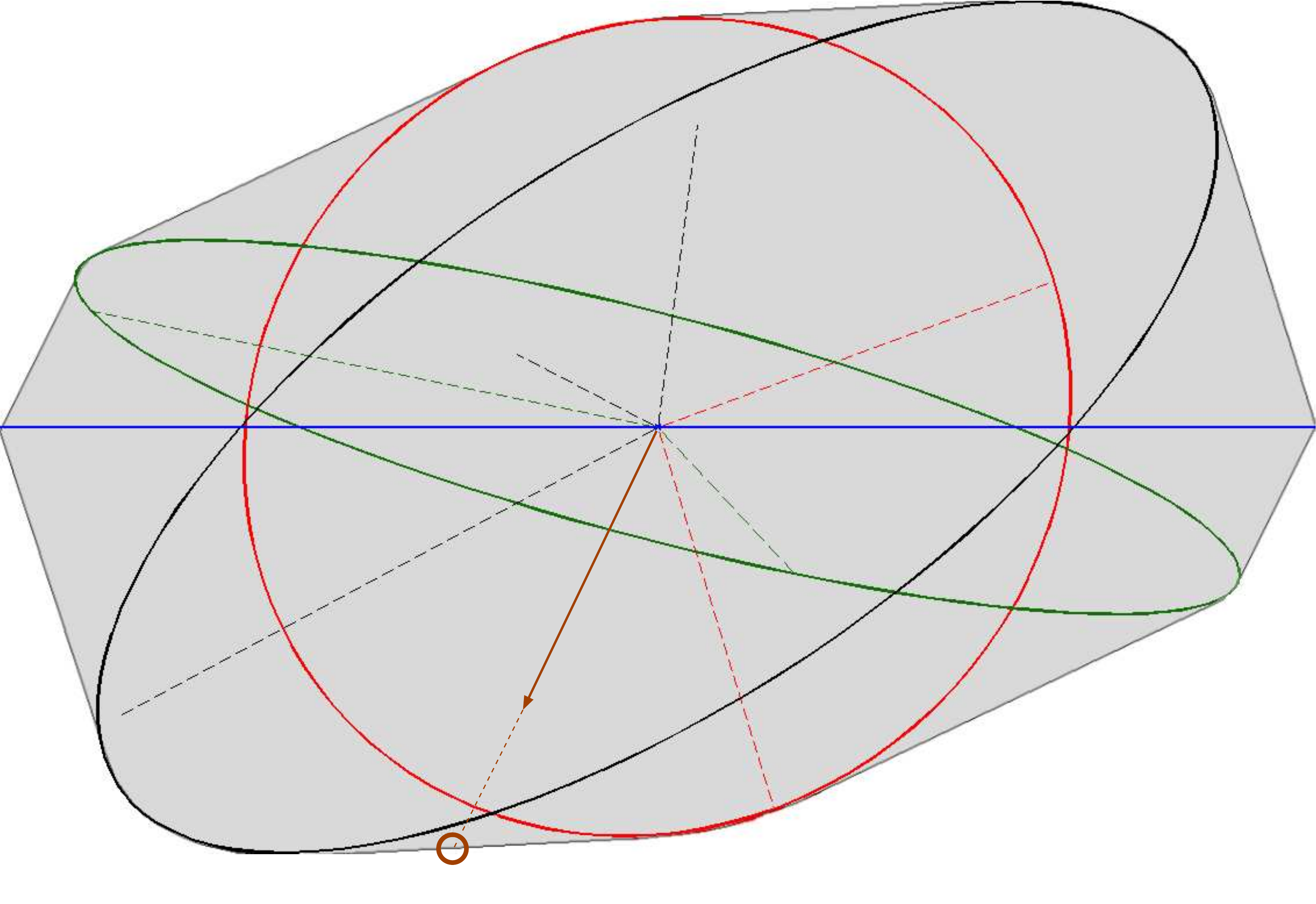}%
\end{picture}%
\setlength{\unitlength}{1697sp}%
\begingroup\makeatletter\ifx\SetFigFont\undefined%
\gdef\SetFigFont#1#2#3#4#5{%
  \reset@font\fontsize{#1}{#2pt}%
  \fontfamily{#3}\fontseries{#4}\fontshape{#5}%
  \selectfont}%
\fi\endgroup%
\begin{picture}(16980,11653)(1,-10814)
\put(13651,-2761){\makebox(0,0)[lb]{\smash{{\SetFigFont{14}{16.8}{\familydefault}{\mddefault}{\updefault}{\color[rgb]{0,0,0}$\textcolor{red}{\vec{a}_{11}}$}%
}}}}
\put(10076,-9936){\makebox(0,0)[lb]{\smash{{\SetFigFont{14}{16.8}{\familydefault}{\mddefault}{\updefault}{\color[rgb]{0,0,0}$\textcolor{red}{\vec{a}_{12}}$}%
}}}}
\put(10051,-6886){\makebox(0,0)[lb]{\smash{{\SetFigFont{14}{16.8}{\familydefault}{\mddefault}{\updefault}{\color[rgb]{0,0,0}$\textcolor{darkgreen}{\vec{a}_{21}}$}%
}}}}
\put(9151,-886){\makebox(0,0)[lb]{\smash{{\SetFigFont{14}{16.8}{\familydefault}{\mddefault}{\updefault}{\color[rgb]{0,0,0}$\vec{a}_{31}$}%
}}}}
\put(1126,-2911){\makebox(0,0)[lb]{\smash{{\SetFigFont{14}{16.8}{\familydefault}{\mddefault}{\updefault}{\color[rgb]{0,0,0}$\textcolor{darkgreen}{\vec{a}_{22}}$}%
}}}}
\put(6451,-3511){\makebox(0,0)[lb]{\smash{{\SetFigFont{14}{16.8}{\familydefault}{\mddefault}{\updefault}{\color[rgb]{0,0,0}$\vec{a}_{32}$}%
}}}}
\put(1501,-8686){\makebox(0,0)[lb]{\smash{{\SetFigFont{14}{16.8}{\familydefault}{\mddefault}{\updefault}{\color[rgb]{0,0,0}$\vec{a}_{33}$}%
}}}}
\put(16076,-4996){\makebox(0,0)[lb]{\smash{{\SetFigFont{14}{16.8}{\familydefault}{\mddefault}{\updefault}{\color[rgb]{0,0,0}$\textcolor{blue}{\vec{a}_{41}}$}%
}}}}
\put(13426, 14){\makebox(0,0)[lb]{\smash{{\SetFigFont{25}{30.0}{\familydefault}{\mddefault}{\updefault}{\color[rgb]{0,0,0}$\mathcal{E}_3$}%
}}}}
\put(14776,-6661){\makebox(0,0)[lb]{\smash{{\SetFigFont{25}{30.0}{\familydefault}{\mddefault}{\updefault}{\color[rgb]{0,0,0}$\textcolor{darkgreen}{\mathcal{E}_2}$}%
}}}}
\put(6151,-761){\makebox(0,0)[lb]{\smash{{\SetFigFont{25}{30.0}{\familydefault}{\mddefault}{\updefault}{\color[rgb]{0,0,0}$\textcolor{red}{\mathcal{E}_1}$}%
}}}}
\put(376,-5511){\makebox(0,0)[lb]{\smash{{\SetFigFont{25}{30.0}{\familydefault}{\mddefault}{\updefault}{\color[rgb]{0,0,0}$\textcolor{blue}{\mathcal{E}_4}$}%
}}}}
\put(5701,-11011){\makebox(0,0)[lb]{\smash{{\SetFigFont{25}{30.0}{\familydefault}{\mddefault}{\updefault}{\color[rgb]{0,0,0}$\textcolor{hellbrown}{t^* \vec{c}}$}%
}}}}
\put(7051,-8086){\makebox(0,0)[lb]{\smash{{\SetFigFont{25}{30.0}{\familydefault}{\mddefault}{\updefault}{\color[rgb]{0,0,0}$\textcolor{hellbrown}{\vec{c}}$}%
}}}}
\put(8276,-10350){\makebox(0,0)[lb]{\smash{{\SetFigFont{14}{30.0}{\familydefault}{\mddefault}{\updefault}{\color[rgb]{0,0,0}$\textcolor{red}{\vec{x_1}}$}%
}}}}
\put(3376,-10500){\makebox(0,0)[lb]{\smash{{\SetFigFont{14}{30.0}{\familydefault}{\mddefault}{\updefault}{\color[rgb]{0,0,0}$\vec{x_3}$}%
}}}}
\end{picture}%
%------------------------------------------------
% End of Xfig Input
%\includegraphics[width=\textwidth]{cOptMultiPDF.pdf}
\end{center}
\caption{In the multiresponse case, the generalized Elfving set $\overline{\mathcal{E}}$ is the convex hull of the ellipsoids $\mathcal{E}_i$.
On this picture, we have plotted the rows of the observation matrices: $\vec{a}_{ij}^T$ is the $j^\textrm{th}$ row
of $A_i$. In the (common) case where $l\leq m$, the vectors $(\vec{a}_{ij})_{j\in[l_i]}$
are on the boundary of the ellipsoid $\mathcal{E}_i$ (here, this is the case for
$\mathcal{E}_1,\mathcal{E}_2,$ and $\mathcal{E}_4$, but not for $\mathcal{E}_3$ since $l_3=3>2$).
Also note that when $l<m$, the ellipsoid $\mathcal{E}_i$
is not full dimensional (on the picture, we have $l_4=1<2$, such that $\mathcal{E}_4$ is a segment).
The intersection of the line directed by $\vec{c}$ and the generalized Elfving set (denoted by a brown circle on the
figure) indicates the weights of the $\vec{c}-$optimal design. Here, $t^*\vec{c}$ is at equal distance
of the two extremal points $\vec{x}_1 \in \mathcal{E}_1$ and $\vec{x}_3 \in \mathcal{E}_3$, so that
the $\vec{c}-$optimal design is $\vec{w}=[0.5,0,0.5,0]^T$. \label{fig:ElfvingExt}}
\end{figure}

\begin{remark}
The latter theorem has a simple geometric interpretation. In the scalar case, we have seen that
the $\vec{c}-$optimal design could be find at the intersection of a polyhedron and a straight
line directed by $\vec{c}$ (see Figure~\ref{fig:Elfving}). In the multiresponse case, the generalized Elfving set is no longer a polyhedron: instead, we compute the intersection between the straight line directed by $\vec{c}$ and the set
\begin{align*}
\mathcal{\overline{E}}&=\operatorname{convex-hull} \big\{ A_i^T \vec{\epsilon_i},\ i\in[s],\
\vec{\epsilon_i} \in \R^{l},\ \Vert \vec{\epsilon_i} \Vert \leq 1 \big\}, \\
&=\operatorname{convex-hull} \big\{ \mathcal{E}_i,\ i \in[s] \big\},
\end{align*}
where $\mathcal{E}_i$ is the ellipsoid with semi-axis $\sqrt{\lambda_k^{(i)}} \vec{u_k^{(i)}}$ ($k \in [m]$), where 
$\{\lambda_1^{(i)},\ldots,\lambda_m^{(i)}\}$ are the eigenvalues of $A_i^T A_i$ and
$\{\vec{u_1^{(i)}},\ldots,\vec{u_m^{(i)}}\}$ are the corresponding eigenvectors. In the common case,
we have $l<m$, such that some eigenvalues of $A_i^T A_i$ vanish and the ellipsoid $\mathcal{E}_i$
is not full-dimensional (i.e.\ its volume is zero). We illustrate this
geometric interpretation in Figure~\ref{fig:ElfvingExt}.
Moreover, we see in next theorem that the intersection which characterizes the $\vec{c}-$optimality
can be computed by a Second order cone program.
\end{remark}

\begin{theorem}[Computation of the $\vec{c}-$optimal design by SOCP]
\label{theo:SOCP}
Assume that the number of available experiments is $s$,
so that $\mathcal{X}$ can be identified with $[s]$.
Let $\vec{u^*},(\vec{\mu^*},\vec{h_i^*})$ be a pair
of primal and dual solutions of the second order cone programs:

\begin{align}
 \textrm{(P-SOCP)}:\ &\max_{\vec{u} \in \R^m} \quad \vec{c}^T \vec{u}  \label{P-SOCP}\\
&\qquad \forall i \in [s], \quad \Vert A_i \vec{u} \Vert \leq 1 \nonumber
~\nonumber \\
\textrm{(D-SOCP)}:\ &\min_{\substack{\vec{\mu} \in \R^s\\ \vec{h_i} \in \R^{l}}} \quad \sum_i \mu_i \label{D-SOCP}\\
&\qquad \vec{c}=\sum_{i=1}^s A_i^T \vec{h_i}\nonumber \\
&\qquad \forall i \in [s],\quad  \Vert \vec{h_i} \Vert \leq \mu_i. \nonumber
\end{align}

We define
\begin{equation*}
\vec{w} := t \vec{\mu^*}, \qquad \textrm{where} \qquad t=(\sum_{i=1}^s \mu^*_i)^{-1}.
\end{equation*}
%that is, $\vec{w}$ is a normalization of the vector $\vec{\mu^*}$.
Then $\vec{w}$ is a $\vec{c}-$optimal design. Moreover,
$\hat{\zeta}=\sum \vec{h_i^*}^T \vec{y_i}$ is the best linear estimator of $\vec{c}^T \vec{\theta}$,
and the optimal variance is $\operatorname{var}(\hat{\zeta})
=t^{-2}=(\sum_i \mu_i^*)^2=(\vec{c}^T \vec{u^*})^2$.
\end{theorem}

\begin{proof}
This result is actually a corollary of Theorem~\ref{theo:ElfvingExt}.
As in the proof of the latter theorem, define $t$ as the largest scalar such that
 $t\vec{c} \in \overline{\mathcal{E}}$, i.e.\ such that
there exists $w_i$ summing to $1$ and vectors $\vec{\epsilon_i}$ in the unit ball of $\R^l$ satisfying
$$t\vec{c}=\sum_{i=1}^s w_i A_i^T \vec{\epsilon_i}.$$ This decomposition gives the optimal weights $w_i$ and the best estimator of $\zeta$:
\begin{equation}
\hat{\zeta}=\sum_{i=1}^s \vec{h_i}^T \vec{y_i}, \label{bestEstimateInProof} 
\end{equation}
where $\vec{h_i}=\frac{w_i}{t} \vec{\epsilon_i}$. According to the proof of Theorem~\ref{theo:ElfvingExt}
indeed, an unbiased estimator of the form~\eqref{bestEstimateInProof} is optimal if and only if every
$\vec{h_i}$ is proportional to $\vec{\epsilon_i}$ and has norm $\frac{w_i}{t}$.
Setting $\vec{z_i}=w_i \vec{\epsilon_i}$, one obtains
$t$ as the value of the following SOCP:
\begin{align}
 \max_{t,\vec{z_i},\vec{w}}\ & \quad t \label{cOptSOCPt}\\
s.t.\quad & \quad t\vec{c}=\sum_{i=1}^s A_i^T \vec{z_i}, \nonumber\\
 & \quad \forall i \in [s],\quad \Vert \vec{z_i} \Vert \leq w_i,\nonumber\\
 & \quad \sum_i w_i=1,\quad \vec{w}\geq\vec{0} \nonumber.
\end{align}
In order to get an SOCP in the standard form, we write $w_i=t \vec{\mu_i}$,
where $t=\frac{1}{\sum_i \mu_i}$ is an arbitrary nonnegative scalar.
Then, we set $\vec{h_i}=t^{-1}\vec{z_i}$, and we obtain a problem in the form of~\eqref{D-SOCP}. Finally, the value of $(P-SOCP)$ and $(D-SOCP)$ are equal,
since the Slater condition holds for this pair of programs (the dual $(D-SOCP)$ is strictly feasible and
the primal $(P-SOCP)$ is feasible).
A proof of the strong duality theorem for SOCP can be found e.g.\
in~\cite{NN94}, Section 4.2. See~\cite{LVBL98} for more background
on SOCP duality theory.
\end{proof}

The previous theorem shows how one can compute the $\vec{c}-$optimal design on a finite regression region by solving an SOCP.
This can be done with the help of interior points codes such as SeDuMi~\cite{sedumi}.
Solving the latter SOCP~\eqref{P-SOCP} is a much easier task than solving 
the former state-of-the-art SDP for $\vec{c}-$optimality,
because the number of variables is in the order of $m$ (instead of $m^2$); because we have get
rid off the positive semidefiniteness constraint of the SDP; and because the SOCP solver is
able to exploit the sparse structure of the observation matrices $A_i$ (while the partial information
matrices $M_i=A_i^TA_i$ involved in the SDP are \emph{not very sparse} in general.
As stated in the introduction,
this approach has been used for the computation of $\vec{c}-$optimal
designs for a network application~\cite{SagnolGB10ITC}. Very large instances of experimental design problems
arise for the optimal monitoring of IP networks indeed, and we will see in Section~\ref{sec:numexpNetflow}
that the present SOCP can handle them.

\section{Two extensions of the main result}

In this section, we show that optimal designs can be computed by SOCP in two particular situations
which generalize $\vec{c}-$optimality,
namely when the experimenter wants to estimate several quantities $\vec{c_1}^T\vec{\theta},\ldots,\vec{c_r}^T\vec{\theta}$ and
chooses the $A-$optimality criterion, and when the design is subject to multiple resource constraints.

\subsection{The case of $A-$optimality}
\label{sec:Ades}
When there are several quantities of interest, i.e.\ when 
$\vec{\zeta}=(\vec{c_1}^T\vec{\theta},\ldots,\vec{c_r}^T\vec{\theta})^T$ consists in a collection
of $r$ linear combinations of the parameters ($\vec{\zeta}=K^T \vec{\theta},$
where $K$ is the $m \times r$ matrix formed by the columns $\vec{c_i}$), it is known that the best linear unbiased estimator is
$$\vec{\hat{\zeta}}= K^T (M(\xi))^\dagger A(\xi)^T \Delta(\vec{w})^{-1} \vec{y}(\xi),$$ and its covariance matrix is
$$\operatorname{Var}(\vec{\hat{\zeta}})=K^T (M(\xi))^- K.$$
Notice that an interesting case occurs when $K=\vec{I}$, i.e.\ when the experimenter wants to estimate the whole vector of parameters.
As in the case of $\vec{c}-$optimality, we say that $\vec{\zeta}=K^T \vec{\theta}$ is estimable if there is a design $\xi$ such that
$\range K \subset \range A(\xi)$, and we denote by $\Xi(K)$ the feasibility cone (i.e.\ the set of designs $\xi$ satisfying the latter range inclusion).

The objective is now to \textit{minimize}, in a certain sense, the variance of this best estimator. A widely used criterion
is to minimize the trace of this matrix : such a minimizing design is known as $A-$optimal. 
\begin{align} \min_{\xi\in\Xi(K)}\ \ & \operatorname{trace}(K^T M(\xi)^- K) \label{Aopt}\\
\operatorname{s.t.}\quad &  M(\xi)=\sum_{i=1}^s w_i A(\vec{x_i})^T A(\vec{x_i}) \nonumber\\
&  \sum_{i=1}^s w_i=1;\quad \forall\ i \in [s],\ w_i\geq 0,\ \vec{x_i}\in \mathcal{X}. \nonumber
\end{align}

We show in this section that computing the $A-$optimal design for the parameter of interest $K^T \vec{\theta}$ can be written as
a $\vec{c}-$optimal design problem with multidimensional observations. 
The objective function of~\eqref{Aopt} can indeed be written as
$$\trace\ (K^T M(\xi)^- K) = \sum_{k=1}^r \vec{c_k}^T M(\xi)^- \vec{c_k}.$$
We define the vector $\vec{\tilde{c}}$ as the vertical
concatenation of the columns $\vec{c_i}$, i.e.\ $\vec{\tilde{c}}=[\vec{c_1}^T,...,\vec{c_r}^T]^T$. Now , we have:
$\operatorname{trace}(K^T M(\xi)^- K) = \vec{\tilde{c}}^T \tilde{M}(\xi)^- \vec{\tilde{c}},$ where:
\begin{align*}\tilde{M}(\xi)=\left( \begin{array}{ccc}
		M(\xi) & &\\
			& \ddots & \\
			& & M(\xi)
                 \end{array} \right) &=\sum_{i=1}^s w_i \left( \begin{array}{ccc}
		A(\vec{x_i})^T A(\vec{x_i}) & &\\
			& \ddots & \\
			& & A(\vec{x_i})^T A(\vec{x_i}) \end{array} \right) \\
&=\sum_{i=1}^s w_i \left( \begin{array}{ccc}
                           A(\vec{x_i}) & &\\
			& \ddots & \\
			& & A(\vec{x_i}) \end{array} \right)^T \underbrace{\left( \begin{array}{ccc}
                           A(\vec{x_i}) & &\\
			& \ddots & \\
			& & A(\vec{x_i}) \end{array} \right)}_{\tilde{A}(\vec{x_i})}\\
&=\sum_{i=1}^s w_i \tilde{A}(\vec{x_i})^T \tilde{A}(\vec{x_i}).
\end{align*}

In the latter equation, $\tilde{A}(\vec{x_i})$ contains $r$ blocks and is of dimension $ rl \times rm$.
We can now rewrite Problem~\eqref{Aopt} 
in the following form:
% as:
\begin{align*} \min_\xi\ &\quad \trace\ (\vec{\tilde{c}}^T \tilde{M}(\xi)^- \vec{\tilde{c}}) \\
\operatorname{s.t.}\ &\quad \tilde{M}(\xi)=\sum_{i=1}^s w_i \tilde{A}(\vec{x_i})^T \tilde{A}(\vec{x_i}), \\
&\quad  \sum_{i=1}^s w_i=1;\quad \forall\ i \in [s],\ w_i\geq 0,\ \vec{x_i}\in \mathcal{X}. \nonumber
\end{align*}
We have thus shown that the problem of finding the $A-$optimal design is nothing but a
$\vec{\tilde{c}}-$optimal design problem, with
augmented observation matrices $\tilde{A}(\vec{x_i})$. As a consequence,
the results of Section~\ref{sec:multiresponse} on $\vec{c}-$optimality also apply for the
more general $A-$optimal design problem for a subsystem $K^T \vec{\theta}$
of the parameters. In particular, the $A-$optimal design problem reduces to an SOCP:

\begin{theorem}[Computation of the $A-$optimal design by SOCP]
\label{theo:ASOCP}
Let $\big( U^*,(\vec{\mu^*},(H_i^*)_{i\in[s]}) \big)$ be a pair
of primal and dual solutions of the second order cone programs:

\begin{align}
\max_{U \in \R^{m \times r}} &\quad \trace\ K^T U  \label{AP-SOCP}\\
&\quad \forall i \in [s], \quad \Vert A_i U \Vert_F \leq 1 \nonumber\\
& ~\nonumber \\
\min_{\substack{\vec{\mu} \in \R^s\\ H_i \in \R^{l \times r}}} &\quad \sum_i \mu_i \label{AD-SOCP}\\
&\quad K=\sum_{i} A_i^T H_i \nonumber \\
&\quad \forall i \in [s],\quad  \Vert H_i \Vert_F \leq \mu_i. \nonumber
\end{align}
We define
\begin{equation*}
\vec{w} := t \vec{\mu^*}, \qquad \textrm{where} \qquad t=(\sum_{i=1}^s \mu^*_i)^{-1}.
\end{equation*}
%that is, $\vec{w}$ is a normalization of the vector $\vec{\mu^*}$.
Then, $\vec{w}$ is $A-$optimal for $K^T\vec{\theta}$. Moreover,
$\vec{\hat{\zeta}}=\sum_i (H_i^*)^T \vec{y_i}$ is the best linear unbiased estimator of $K^T \vec{\theta}$,
and the optimal $A-$criterion is
$$\sum_{i=1}^r \vec{c_i}^T M(\vec{w}^*)^- \vec{c_i} =t^{-2}=(\sum_i \mu_i^*)^2=( \trace\ K^T U^*)^2.$$
\end{theorem}

We further show that the geometrical characterization of $\vec{c}-$optimality for multiresponse experiments
generalizes the result of Studden~\cite{Stud71}, who established an Elfving-type result for
$A-$optimal designs of single-response experiments ($l=1$ and $A(\vec{x})=\vec{a_x}^T$ is a row vector).
This characterization is based on
the following extension of the Elfving set when the matrix $K$ is $m \times r$:
$$\mathcal{E}_S=\operatorname{convex-hull}\{ \vec{a_x} \vec{\epsilon}^T | \vec{x}\in\mathcal{X},\
\vec{\epsilon} \in \R^r,\ \Vert \vec{\epsilon} \Vert \leq 1 \} \subset \R^{m \times r}$$

\begin{theorem}[Studden,1971]
A design $\xi=\{\vec{x_i},w_i\}$ is $A-$optimal for $K^T \vec{\theta}$ if and only if there exists a scalar $t>0$ and 
vectors $\vec{\epsilon_i}$ in the unit ball of $\R^r$ such that
$$t K = \sum_i w_i \vec{a_{x_i}} \vec{\epsilon_i}^T \in \partial \mathcal{E}_S.$$
Moreover, $t^{-2}=\operatorname{trace} (K^T M(\xi)^- K)$ is the optimal value of the $A-$criterion.
\end{theorem}

One can easily verify that this theorem is a particular case of Theorem~\ref{theo:ElfvingExt}. Using the previously introduced notation indeed,
Theorem~\ref{theo:ElfvingExt} says that $\xi=\{\vec{x_i},w_i\}$ is $A-$optimal for $K^T \vec{\theta}$ if and only if there exists a scalar $t>0$ and vectors
$\vec{\epsilon_i}$ in the unit ball of $\R^{rl}$ such that
$$t \vec{\tilde{c}} = \sum_i w_i \tilde{A}(\vec{x_i})^T \vec{\epsilon_i} \in \partial \mathcal{\overline{E}},$$
and we notice that $\vec{\tilde{c}}$ is the vectorized version of $K$, and when
$l=1$,
$\mathcal{\overline{E}}$ is the vectorized version of $\mathcal{E}_S$ and $\tilde{A}(\vec{x_i})^T \vec{\epsilon_i}=
[\epsilon_{i1} \vec{a_{x_i}}^T,\ldots,\epsilon_{is} \vec{a_{x_i}}^T]^T$ is the vectorized version of
$\vec{a_{x_i}} \vec{\epsilon_i}^T$.

\subsection{Optimal designs subject to multiple resource constraints}
\label{sec:multCons}
The great advantage of the mathematical programming formulations (LP,SOCP,SDP,...) resides mostly in
their flexibility, and the possibility to add ``without effort'' new constraints in the problem. 
Elfving studied the case in which the available experiments have different costs~\cite{Elf52}. If the cost of the $i\th$ experiment
is $p_i$, and the experimenter disposes of a budget $b$, the constraint becomes:
$$\sum_{i=1}^s w_i p_i \leq b.$$ Now, $w_i$ can not be interpreted as the \emph{percentage of experimental effort to spend on the $i\th$
experiment} anymore. Instead, the quantity $w_i \frac{p_i}{b}$ should be seen as the percentage of budget to allocate to the experiment $i$.
Elfving noticed that the change of variable $w_i'=w_i \frac{p_i}{b}$ brings the problem back to the standard situation, and is equivalent
to a scaling of the observation equations~\eqref{obsequations}.

Consider now the more general case in which $\vec{w}$ is a control variable for the experiments, such that the information matrix takes the
standard form $M(\vec{w})=\sum_{i=1}^s w_i A_i^T A_i$ for some observation matrices $A_i$. We assume that $\vec{w}$ is constrained by several
linear inequalities
\begin{equation}
 R \vec{w} \leq \vec{b}, \label{ineqRwb}
\end{equation}
where $\vec{b}\in\R^n,$ $R$ is a $n\times s$ matrix and the inequality is elementwise. Contrarily to the previous situation with
a single \emph{budget constraint}, there is no simple change of variable which brings the problem back to the standard case ($\sum_i w_i=1$),
because we do not know which inequalities will be saturated in~\eqref{ineqRwb} at optimality. This
constrained problem has been studied by Cook and Fedorov~\cite{CF95}, who proposed an extension of
the Fedorov exchange algorithm. However, the latter
exhibits a slow convergence in practice.

This constrained framework arises in the problem of optimally setting the sampling rates of a measuring device on a network.
In this case, $\vec{w}$
is the vector of the sampling rates of the monitoring tool at different locations of the network, and the constraint $R \vec{w} \leq \vec{b}$
reflects the fact that only a certain number of packets should be sampled at each router~\cite{SagnolGB10ITC}.
Another example of application of optimal design problems with multiple constraints was given by
Vandenberghe, Boyd and Wu~\cite{VBW98}: they have shown that constraints of the form~\eqref{ineqRwb}
can be used to avoid \emph{concentrated} designs. For example, we can impose that no more than a given fraction
of the experimental effort, say $90\%$, is concentrated
on a small number of experiments, say $10\%$ of the possible observations. 

Adding those \emph{multiple resource constraints} in the SDP formulation of the
$\vec{c}-$optimal design problem is straightforward. Doing the same thing for the SOCP
formulation is a little more tricky, since some \emph{hyperbolic constraints} need be
reformulated as conic constraints. We do this in the next theorem. 
A related result was obtained by Ben-Tal and Nemirovskii~\cite{BTN92}, for an application
to truss topology design (see also~\cite{NN94,LVBL98}). Our statement of the theorem shows that
the optimal variables of the SOCP moreover give
the best estimator of $\vec{c}^T\vec{\theta},$ and  that the result holds
even if the optimal information matrix is singular (but contains $\vec{c}$ in its range).

\begin{theorem}
The following SOCP is feasible if and only if
$\vec{c}^T\vec{\theta}$ is estimable for a feasible design
(i.e.\ $\exists \vec{w}\geq\vec{0}:\ R\vec{w}\leq\vec{b}$ and $\vec{c} \in \range M(\vec{w})$).
\begin{align}
 \min_{\substack{\vec{w}\geq{0},\ \vec{\mu}\geq\vec{0}\\ \vec{h_i}\in\R^l}} &\quad \sum_{i=1}^s \mu_i \label{SOCP-gen}\\
&\quad \sum_{i=1}^s A_i^T \vec{h_i}=\vec{c} \nonumber\\
&\quad R \vec{w} \leq \vec{b},\ \vec{w}\geq\vec{0} \nonumber\\
&\quad \left\Vert \left[ \begin{array}{c} 2 \vec{h_i} \\ w_i-\mu_i \end{array} \right] \right\Vert \leq w_i+\mu_i,\ (i=1,\ldots,s). \nonumber
\end{align}
If the latter SOCP moreover admits a solution $(\vec{w^*},\vec{\mu^*},(\vec{h_i^*})_{i\in[s]})$,
then $\vec{w^*}$ is $\vec{c}-$optimal (in the sense of the general problem with constraints
$R\vec{w}\leq\vec{b}$), the best unbiased linear estimator of $\zeta=\vec{c}^T \vec{\theta}$ is
$\hat{\zeta}=\sum_i \vec{h_i^*}^T \vec{y_i}$, and the optimal variance is
$\operatorname{var}(\hat{\zeta})=\vec{c}^T M(\vec{w^*})^- \vec{c}=\sum_{i=0}^s \mu_i^*$.
 \end{theorem}

\begin{proof}
Let $\vec{w}$ be a feasible design ($R\vec{w}\leq\vec{b},\ \vec{c}\in\range M(\vec{w})$).
The Gauss Markov Theorem allows us to rewrite the objective criterion of
the $\vec{c}-$optimal design problem as:
\begin{align}
\vec{c}^T M(\vec{w})^- \vec{c}=\min_{\vec{h}\in\R^{sl}} &\quad \vec{h}^T \Delta(\vec{w}) \vec{h}\\
\operatorname{s.t.} &\quad [A_1^T,\ldots,A_s^T]\vec{h}=\vec{c}, \nonumber
\end{align}
where $\Delta(\vec{w})$ is defined in Equation~\eqref{Deltaw}, and the optimal vector $\vec{h}$ in this
problem defines the
best linear unbiased estimator $\hat{\zeta}=\vec{h}^T\vec{y(\xi)}$ of $\zeta=\vec{c}^T \vec{\theta}$.
Decomposing $\vec{h}$ as $[\vec{h_1}^T,\ldots,\vec{h_s}^T]^T,$  $\vec{h_i}\in \R^{l}$,
the expression $\vec{h}^T \Delta(\vec{w}) \vec{h}$ can be rewritten as
\begin{equation}
\sum_{i=1}^s w_i^{-1} \Vert \vec{h_i} \Vert^2. \label{sumfrachw}
\end{equation}
Recall that when an experiment is unobserved ($w_i=0$), it could simply be removed from the support of
the experimental design. In other words, the sum~\eqref{sumfrachw} is taken on the indices such
that $w_i>0$ only. We can now rewrite the $\vec{c}-$optimal design problem with constraints $R\vec{w}\leq\vec{b}$
in a form that involves the vector of coefficients $\vec{h}$ of the estimator $\hat{\zeta}$:
\begin{align}
\min_{\vec{w},\ (\vec{h_i}\in\R^{l})_{i\in[s]}} &\quad 
\sum_{\{i: w_i>0\}} \frac{\Vert \vec{h_i} \Vert^2}{w_i}\\
\operatorname{s.t.}\quad &\quad \sum_{i=1}^s A_i^T\vec{h_i}=\vec{c}, \nonumber\\
&\quad R\vec{w}\leq \vec{b},\ \vec{w}\geq \vec{0}. \nonumber
\end{align}
Clearly, this is equivalent to:
\begin{align}
\min_{\substack{\vec{w}\geq\vec{0},\ \vec{\mu}\geq\vec{0}\\ \vec{h_i}\in\R^{l}}} &\quad
\sum_{i=1}^s \mu_i \label{SOCP-hyp}\\
\operatorname{s.t.}\quad &\quad \sum_{i=1}^s A_i^T\vec{h_i}=\vec{c}, \nonumber\\
&\quad R\vec{w}\leq \vec{b}, \nonumber\\
&\quad \Vert \vec{h_i} \Vert^2 \leq \mu_i w_i,  \nonumber
\end{align}
since we can assume without loss of generality that $w_i=0 \Rightarrow \Vert \vec{h_i} \Vert = \mu_i=0$.
Finally, the SOCP~\eqref{SOCP-gen} is obtained by reformulating the hyperbolic constraints
$\Vert \vec{z} \Vert^2 \leq \alpha \beta$ as
$$\left\Vert \vectwo{2\vec{z}}{\alpha-\beta} \right\Vert \leq \alpha+\beta.$$

We now show that Problem~\eqref{SOCP-hyp} is feasible if and only if the quantity $\vec{c}^T\vec{\theta}$
is estimable for a feasible design. We first assume the contrary, i.e.\ $R\vec{w}\leq\vec{b},\ \vec{w}\geq\vec{0}
\Rightarrow \vec{c}\notin\range\ M(\vec{w})$. Let $\vec{w}$ satisfies the inequality constraints; we denote by $\mathcal{I}=\{i_1,\ldots,i_p\}$
the subset of indices such that $w_i>0$, so that we have
$$\range\ M(\vec{w})=\range\ [A_{i_1}^T,\ldots,A_{i_p}^T].$$
Now, if $\sum_{i=1}^s A_i^T\vec{h_i}=\vec{c}$ for some vectors $\vec{h_i}$, it implies that $\vec{h_i}\neq\vec{0}$ for an index
$i$ such that $w_i=0$, and the hyperbolic constraint $ \Vert \vec{h_i} \Vert^2 \leq \mu_i w_i$ is violated.
Conversely, let $\vec{w}$ be a feasible design, and $\mathcal{I}=\{i_1,\ldots,i_p\}$ defined as above. Since  $\vec{c}\in\range\ M(\vec{w}),$
we can write $\sum_{k=1}^p A_{i_k}^T\vec{h_{i_k}}=\vec{c}$ for some vectors $\vec{h_{i_1}},\ldots,\vec{h_{i_p}}$. Finally, we define
$\mu_i=\frac{\Vert \vec{h_i} \Vert^2}{w_i}$ for all $i\in\mathcal{I}$, and
$\vec{h_j}=\vec{0},\ \mu_j=0$ for all the other indices $j\in[s]\setminus\mathcal{I}$: Problem~\eqref{SOCP-hyp} is feasible
for the vectors $\vec{w},\vec{\mu}$ and $(\vec{h_i})_{i\in[s]}$.
However, we point out that Problem~\eqref{SOCP-hyp} (as well as the $\vec{c}-$optimal design problem
with multiple constraints) may fail to have a solution when the polyhedron $\{\vec{w}\geq\vec{0}: R\vec{w}\leq\vec{b}\}$ is not bounded, because in some cases the minimum is not attained and the infimum
can be approached by a sequence of designs $(\vec{w_t})_{t\in \mathbb{N}}$ such that $\Vert \vec{w_t} \Vert \underset{t\to+\infty}{\longrightarrow} \infty$.
\end{proof}

\section{Alternative approach via Lagrange duality}
\label{sec:lagrange}
The aim of this section is to prove directly the equivalence between Problem~\eqref{mincMc} and the Second
Order Cone Programs~\eqref{P-SOCP}-\eqref{D-SOCP}
by use of Lagrangian duality tools, when the number of available experiments is finite (or equivalently that the locations of the selected
experiments are given in advance). As stated in the introduction, the motivation for this work is to understand why the former SDP approach of this problem actually reduces to a much easier problem.
%the relation between this convex conic optimization problem with the original nonconvex problem of statistics, and to some extents,
%to make the community of optimizers aware of the nice properties of this problem, which is mostly studied by statisticians.
Moreover, our proof handles the multiresponse case in the same manner as the single-response case, by simply 
substituting observation row vectors with observation matrices
in the constraints of the optimization problem, and it can be applied in a more general framework (cf.\
Section~\ref{sec:Sopt}).

We first recall that the $\vec{c}-$optimal design problem~\eqref{mincMc} is equivalent to a semidefinite
program. The $\vec{c}-$optimality SDP already appeared in Pukelsheim~\cite{Puk80}, hidden under a more general
form. The proof given in this paper is more elementary, and it can be adapted to handle $S-$optimality
(cf.\ Lemma~\ref{lem:P/Dbeta-SDP}). In the sequel, $\succeq$ denotes the L\"owner ordering of
symmetric matrices, that is, $B \succeq C$ if and only if $B-C$ is positive semidefinite.

\begin{theorem}
\label{theo:SDP}
When the regression range is finite ($\mathcal{X}=[s]$), the $\vec{c-}$optimal design problem~\eqref{mincMc}
is equivalent to the following pair of primal and dual SDP. More precisely, let $(X^*,\vec{\mu^*})$ be a pair of primal and dual solution of the  SDPs:
\begin{align}
(P-SDP)\qquad\qquad \max_{X \succeq 0} & \quad \vec{c}^T X \vec{c} \label{maxcNc} \\
\textrm{s.t.} & \quad \mathrm{trace }(A_i X A_i^T) \leq 1,\indent \forall i \in [s] \nonumber\\
~\nonumber\\
(D-SDP)\qquad\qquad \min_{\vec{\mu}\geq\vec{0}} & \quad \sum_i \mu_i  \label{DSDP}\\
\textrm{s.t.} & \quad \sum_i \mu_i  A_i^T A_i \succeq \vec{cc}^T, \nonumber
\end{align}
Then, the vector $\vec{w}=\frac{\vec{\mu^*}}{\sum_i \mu_i^*}$ is solution of the $\vec{c}-$optimal design problem~\eqref{mincMc}, and the following relation holds:
$$\vec{c}^TM(\vec{w})^-\vec{c}=\sum \mu_i^*=\vec{c}^T X^* \vec{c}.$$
\end{theorem}

\begin{proof}
Since the feasible designs $\vec{w}$ must be such that $\vec{c}\in \range M(\vec{w})$, we can write, using a generalized Schur complement:
$$\left\lbrace \begin{array}{c}
   t\geq \vec{c}^T M(\vec{w})^- \vec{c}\\
   t\geq0
  \end{array} \right.
 \Longleftrightarrow \left( \begin{array} {c|c} M(\vec{w}) & \vec{c}\\
                       			\hline
					 \vec{c}^{{}_T} & t
                      \end{array} \right) \succeq 0,$$
for $t \in \R$. Note that if we exclude the trivial case $\vec{c}=\vec{0}$, we can always assume that $t>0$.
%(If $t=0$, then there is a design $\vec{w}$ for which
% $\vec{c}^T M(\vec{w})^- \vec{c}=0,$ and since $\vec{c}\in \range M(\vec{w})$, this implies
% $\vec{c}=\vec{0}$).
 So, the $\vec{c}-$optimal design problem~\eqref{mincMc} can be rewritten as:

 \begin{align}
 \min_{t\in \R,\vec{w}\geq\vec{0}} &\quad t \label{SDPackingDual2}\\
 \textrm{s.t.} &\quad \left( \begin{array} {c|c} M(\vec{w}) & \vec{c}\\
                       			\hline
					 \vec{c}^{{}_T} & t
                      \end{array} \right) \succeq 0. \nonumber \\
&\sum w_i=1. \nonumber
\end{align}

Now, we re-express the positive-semidefiniteness of the matrix in~\eqref{SDPackingDual2} with another Schur complement:
$$\left( \begin{array} {c|c} M(\vec{w}) & \vec{c}\\
                       			\hline
					 \vec{c}^{{}_T} & t
                      \end{array} \right) \succeq 0  \Longleftrightarrow \left\lbrace \begin{array}{c}
   M(\vec{w}) \succeq 0\\
   M(\vec{w}) \geq t^{-1} \vec{c c}^T
  \end{array} \right. .$$
Setting the new variable $\vec{\mu}=t\vec{w}$, so that $t=\sum_i \mu_i$, we obtain a semidefinite program in the form of~$(D-SDP)$. Strong duality
holds between $(D-SDP)$ and $(P-SDP)$, because the Slater condition for semidefinite programming is satisfied (the primal problem $(P-SDP)$ is strictly feasible, and its dual is feasible). Finally, notice that the
dual feasibility condition $\sum_i \mu_i A_i^T A_i \succeq \vec{cc}^T$ implies that $\vec{c}$ is
included in $\range \sum_i \mu_i A_i^T A_i=\range M(\vec{w})$, so that we do not have to worry
about the the constraint $\xi\in\Xi(\vec{c})$ in the initial problem~\eqref{mincMc}.
\end{proof}
\vspace{12pt}
~\\
We notice that when the above SDP has a rank one solution, it can be reduced to the primal
SOCP~\eqref{P-SOCP}.
The next theorem shows that such a rank-one solution always exists indeed,
and its proof can be found in~\cite{Sagnol09rank}. Note that this theorem is of independent interest.
The general result establishes the existence of
rank-$r$ solutions for the class of \emph{semidefinite packing problems} in which
the matrix defining the objective function has rank $r$:

\begin{theorem}[Low rank reduction theorem for semidefinite \emph{packing} problems {{\cite{Sagnol09rank}}}]
\label{theo:SDPSol1}
Denote by $\mathbb{S}_m$ the space of $m \times m$ symmetric matrices, equipped with the inner product $\langle A,B \rangle := \mathrm{trace}(AB^T)$.
We consider the following semidefinite packing program
\begin{align}
 \max_{X \in \mathbb{S}_m} &\ \quad \langle K^T K,X \rangle  \label{stdSDP}\\
s.t. & \quad \langle A_i^T A_i,X \rangle \quad \leq \quad b_i,\quad \forall i \in [s],\nonumber \\
& \quad X \succeq 0, \nonumber
\end{align}
This problem is feasible if and only if
every $b_i$ is nonnegative, and is bounded if and only if
$\range K^T \subset \range (\sum_{i=1}^s A_i^T A_i)$.
Under these two conditions, the SDP~\eqref{stdSDP} has a solution $X$ which is of rank at most
$r:=\operatorname{rank}\ K^T K$.

In particular, if $K=\vec{c}$ is a vector (i.e.\ $\operatorname{rank} K^TK =1$), then there is an optimal
variable in the form $X=\vec{xx}^T$
for some vector $\vec{x}\in \R^m$, and $\vec{x}$ is the solution of the following Second-Order Cone Program:
\begin{align}
 \max_{\vec{x}\in \R^m} &\quad \vec{c}^T \vec{x} \label{SOCPPacking}\\
 \textrm{s.t.} &\quad \Vert A_i \vec{x} \Vert \leq \sqrt{b_i}, \qquad \forall i\in[s].\nonumber
\end{align}
\end{theorem}
\vspace{12pt}

This theorem shows that if $\vec{c}^T\vec{\theta}$ is estimable, then Problem~\eqref{maxcNc} admits a solution of rank-one and reduces to Problem~\eqref{P-SOCP}. This is another proof of Theorem~\eqref{theo:SOCP} which explains why $\vec{c}-$optimal designs can be computed by SOCP. To our sense,
this proof gives a better understanding on how the computing gap from SDP to SOCP is crossed : the intrinsic positive semidefiniteness of
the data in the SDP ensures that there is a solution of rank $1$, which explain why the complexity of
the problem fades. 

\begin{remark}
The rank reduction theorem~\ref{theo:SDPSol1} admits a generalization, which is presented
in~\cite{Sagnol09rank} and allows one to handle the case with multiple constraints $R\vec{w}\leq\vec{b}$.
\end{remark}

\begin{remark}
The theorem above can be extended to a semi-infinite programming context. This allows one to write the $\vec{c}-$optimal design problem over an infinite regression region $\mathcal{X}$ under the form:
\begin{align}
\max_{\vec{u}} & \quad \vec{c}^T \vec{u} \label{SISOCP}\\
&\quad \forall \vec{x} \in \mathcal{X}, \quad \Vert A(\vec{x})\ \vec{u} \Vert \leq 1 \nonumber.
\end{align}
Note that this formulation is given with a theoretic purpose only, since the resolution of this kind of problem is very hard,
even in the case in which $A(\vec{x})$ is linear in $\vec{x}$. If some regularity conditions are satisfied by the application $\vec{x} \mapsto A(\vec{x})$,
then there is a stochastic algorithm which converge to the optimal of Problem~\eqref{SISOCP} with probability $1$~\cite{TMP06}, but
the convergence can be extremely slow. If we can solve this optimization problem, though,
then the points $\vec{x}$ where the inequality constraint is active are nothing but the support of 
the optimal design.
\end{remark}

\section{T-optimality for $K^T \vec{\theta}$}
\label{sec:Topt}
When several quantities are of interest (optimal design for $K^T \vec{\theta}$),
another classic criterion from the experimental design literature is the
$T-$optimal design problem:
\begin{align}
 \sup_{\xi=\{\vec{x_k},w_k\} \in \Xi(K)} &\quad \Phi_T(\xi):= \trace\ (K^T M(\vec{w})^- K)^{-1} \label{Topt}\\
\textrm{s.t.} \qquad &\quad  \sum_{i=1}^s w_i=1;\quad \forall\ i \in [s], w_i\geq 0,\vec{x_i}\in \mathcal{X}. \nonumber
\end{align}
The criterion $\Phi_T$ can be extended
by continuity to the designs $\xi$ such that $K^T M(\vec{w})^- K$ is not invertible,
i.e.\ such that $\xi \notin \Xi(K)$. Contrarily to the $A-$optimality criterion,
there are some examples where the maximum is attained for a nonfeasible design
$\xi\notin\Xi(K)$ (see Pukelsheim~\cite{Puk93}, Section~6.5). Following the
terminology of Pukelsheim, we call a design \emph{formally} $T-$optimal if it maximizes
the objective criterion $\Phi_T$ subject to the constraints $\sum_{i=1}^s w_i = 1$ only,
i.e.\ when we ignore the feasibility condition $\range K \subset \range M(\vec{w})$. Note that
a formally T-optimal design always exists, because the feasibility region becomes compact
when we remove the feasibility constraint.
Pukelsheim shows in Section~{9.15} of~\cite{Puk93} that the $T-$optimal design problem for the full parameter $\vec{\theta}$ is trivial: if $\vec{\theta}$ is estimable ($K=\vec{I}$), then 
a design is formally $T-$optimal for $\vec{\theta}$ if and only if it allocates all the weight to the experiments $i$ such that $\Vert A_i \Vert_F$ is maximal.
However, when the quantity of interest is a parameter subsystem
$\vec{\zeta}=K^T\vec{\theta}$, $K\neq \vec{I}$, the problem becomes computationally challenging.

We show in this section that it is possible to compute a \emph{formally} $T-$optimal design
for $K^T \vec{\theta}$ with a SOCP when $\mathcal{X}$ is finite.
This gives another example of optimal experimental design problem
which can be handled via Second Order Cone Programming.

\begin{theorem}[T-optimality SOCP] \label{theo:Topt}
Let $\mathcal{X}\equiv[s]$ be finite, $K^T \vec{\theta}$ be an estimable quantity, and $\big( (t^*,U^*), (Z_i^*,\vec{w^*},\vec{\gamma^*})\big)$ be a pair of primal and dual solutions of the second order cone programs:
\begin{align}
\min_{t\in\R,\ \vec{U} \in \R^{m \times r}} &\quad t  \label{TP-SOCP}\\
&\quad K^TU=\vec{I}\nonumber\\
&\quad \forall i \in [s], \quad \Vert A_i U \Vert_F^2 \leq t \nonumber\\
& ~\nonumber \\
\max_{\substack{Z_0 \in \R^{r\times r},Z_i \in \R^{l \times r}\\\vec{w}\geq\vec{0},\vec{\gamma}\geq\vec{0}}}
 &\quad -(\trace\ Z_0+\sum_{i=1}^s \gamma_i) \label{TD-SOCP}\\
&\quad KZ_0=\sum_{i=1}^s A_i^T Z_i,\quad \sum_{i=1}^s w_i=1, \nonumber \\
&\quad \forall i \in [s],\quad  \Vert Z_i \Vert_F^2 \leq 4 w_i \gamma_i. \nonumber
\end{align}
(Note that these are Second Order Cone Programs indeed; we have let the
hyperbolic constraints to simplify the notation, otherwise the matrices $A_iU$ and
$Z_i$ need be vectorized). 
Then, $\vec{w^*}$ is formally $T-$optimal for $K^T\vec{\theta}$, and the value of the supremum in Problem~\eqref{Topt} is $t^*=-(\trace(Z_0^*)+\sum_i \gamma_i^*)$.
If moreover $\range K \subset \range M(\vec{w^*})$, then $\vec{w^*}$ is $T-$optimal.
%Otherwise, the supremum in the $T-$optimal design problem~\eqref{Topt} is not attained by any design $\vec{w}\in\Xi(K)$. C'est vrai ?
\end{theorem}

\begin{proof}
Our proof relies on the general definition of the information matrix for $K^T\vec{\theta}$, which is given by Pukelsheim
in Chapter~3 of~\cite{Puk93}:
\begin{align}
C_K(\xi):=&\quad  \min_{U \in \R^{m \times r}} {}_{\preceq}\quad U^T M(\xi) U \label{defQKmin}\\
\operatorname{s.t.} &\quad K^T U=\vec{I}. \nonumber
\end{align}
In the latter expression, the minimum is taken with respect to the L\"owner ordering. Pukelsheim shows that the minimum exists indeed (which
is not obvious since the L\"owner ordering is a partial ordering), as a consequence of the Gauss-Markov
Theorem (cf.\ Theorem~1.21 in~\cite{Puk93}), and moreover that $C_K(\xi)$ coincides with the standard expression $(K^T M(\vec{w})^- K)^{-1}$
for feasible designs $\xi=\{\vec{x_k},w_k\} \in \Xi(K)$.

Now, since the trace of a matrix preserves the L\"owner ordering,
we can express the (formal) $T-$optimal design problem as the following saddle point problem:
\begin{align*}
\max_{\vec{w}\geq\vec{0},\ \sum_i w_i=1}\ \trace C_K(\vec{w})
&=\max_{\vec{w}\geq\vec{0},\ \sum_i w_i=1}\ \min_{U:\ K^T U=\vec{I}_r}\quad  \trace U^T M(\vec{w}) U\\
&=\max_{\vec{w}\geq\vec{0},\ \sum_i w_i=1}\ \min_{U:\ K^T U=\vec{I}_r}\quad \sum_{i=1}^s w_i \Vert A_i U\Vert_F^2\\
&=\min_{U:\ K^T U=\vec{I}_r}\ \left(\max_{i\in[s]}\ \Vert A_i U\Vert_F^2\right).
\end{align*}
The exchange of the max and the min above is a consequence of Sion's minimax theorem
($(\vec{w},U)\mapsto\sum_{i=1}^s w_i \Vert A_i U\Vert_F^2$ is continuous, concave in $\vec{w}$ and convex in $U$).
We next introduce a variable $t$ which must be larger than all the quantities $\Vert A_i U\Vert_F^2$,
and we have shown that the (formal) $T-$optimal
design problem for $K^T\vec{\theta}$ is equivalent to Problem~\eqref{TP-SOCP}.
The (formal) T-optimal design $\vec{w^*}$ is
the optimal dual variable corresponding to the hyperbolic constraints in Problem~\eqref{TP-SOCP}. It follows that
$\vec{w^*}$ can be computed by solving the dual optimization
problem~\eqref{TD-SOCP}. Finally, the value of these optimization problems is the same
by Strong duality (Slater condition holds), and is equal to the optimum of
the $T-$optimal problem~\eqref{Topt}.  

\end{proof}

\section{Convex optimization and $S-$optimality}
\label{sec:Sopt} 

The  $S$-criterion was introduced by L\"auter~\cite{Laut74} in order to tackle the uncertainty of the experimenter on the \emph{true model},
by considering a class of $r$ plausible models with means
$$\mathbb{E}[\vec{y}(\vec{x})]=A_{(k),\vec{x}} \vec{\theta},$$
in which the quantity to estimate is $\zeta_k=\vec{c_k}^T \vec{\theta}\ (\forall k\in[r])$.

In other words, the measurement $\vec{y}(\vec{x})$ at $\vec{x}$ is modeled as a linear function of the parameter $\vec{\theta}$,
which depends on the model, and must be used to estimate a linear function $\zeta$ of the parameter in each model.
In practice, the parameters of each of these models may be different. This can be handled by setting the $j\th$ 
column of $A_{(k),\vec{x}}$ to $\vec{0}$ whenever the $k\th$ model at $\vec{x}$ does not depend on $\theta_j$.
Note  that we write the index of the model in parenthesis, in order to avoid ambiguities with the index of the experiment.

Given a nonnegative vector $\vec{\beta}$ of size $r$ with sum $1$, where $\beta_k$ indicates the importance that the experimenter attaches
to the model $k$, or the importance of the linear combination $\vec{c_k}^T \vec{\theta}$, the $S_{\vec{\beta}}-$criterion is:
$$S_{\vec{\beta}}(\xi)=\sum_{k=1}^r \beta_k \log(\vec{c_k}^T M_{(k)}(\xi)^- \vec{c_k}),$$
where $$M_{(k)}(\xi)=\sum_{i=1}^s w_i A_{(k),\vec{x_i}}^T A_{(k),\vec{x_i}}$$
is the information matrix in the $k\th$ model. A design minimizing this criterion is called $S_{\vec{\beta}}-$optimal.
An interesting case occurs when the $s$ models
are identical. This is an alternative approach to the $A-$optimality for $K^T\vec{\theta}$, with weightings on each linear combination $\vec{c_k}^T \vec{\theta}$ to be estimated.
Dette studied the difference between these two approaches in Section~4 of~\cite{Dette93}.

We are next going to show that the $S_{\vec{\beta}}-$optimal design
of multiresponse experiments
reduces to the problem of maximizing a weighted geometric mean under norm constraints.
This is of great interest for the computation of 
$S_{\vec{\beta}}-$optimal designs. Indeed, this optimization problem is a \emph{geometric program},
and so it can be reformulated in a form for which a self-concordant barrier function is known,
and it can be solved efficiently to the desired precision by interior point techniques~(see e.g.~\cite{BV04}).

\begin{theorem}
\label{theo:SoptMulti}
Let $(\vec{t},(\vec{v_{ik}}),\vec{w})$ be a solution of the following optimization problem. Then,
$\vec{w}$ also minimizes the $S_{\vec{\beta}}-$criterion. Moreover, the value of this program coincides with the value of its dual, which we give below. 
\begin{align}
\min _{\vec{w}\geq\vec{0}, \sum_i w_i=1} S_{\vec{\beta}}(\vec{w}) = & \quad 2 \min_{\vec{t}, (\vec{v_{ik}}), \vec{w}}\ \ \sum_{k=1}^r -\beta_k \log(t_k) \nonumber\\
&\quad t_k \vec{c_k} = \sum_{i=1}^s A_{(k),i}^T \vec{v_{ik}},\qquad \quad \ \forall k \in [r], \nonumber \label{dualVmin} \tag{$P_{\vec{\beta}}$}\\
&\quad \left\Vert \begin{array}{c}
                   \sqrt{\beta_1} \vec{v_{i1}}\\
		   \vdots\\
		   \sqrt{\beta_r} \vec{v_{ir}}
                  \end{array} \right\Vert \leq w_i\qquad \qquad  \forall i\in [s], \nonumber\\
&\quad \sum_{i=1}^s w_i \leq 1. \nonumber\\
=&\quad  2 \max_{\vec{h_1},\ldots,\vec{h_r}}\ \sum_{k=1}^r \beta_k \log \frac{\vec{c_k}^T \vec{h_k}}{\beta_k} \tag{$D_{\vec{\beta}}$} \label{primalVmin}\\
&\quad \left\Vert \begin{array}{c} A_{(1),i} \vec{h_1}/\sqrt{\beta_1}\\
					\vdots \\
					A_{(r),i} \vec{h_r}/\sqrt{\beta_r}          
             \end{array} \right\Vert \leq 1 \qquad \quad \forall i\in[s]. \nonumber
\end{align}
The variables of the
primal optimization problem are $\vec{w}\in \R^m$ (the design), $\vec{t}\in \R^r$ and the vectors $\vec{v_{ik}} \in \R^{l}$, for $i\in[s]$
and $k\in[r]$. The variables of the dual problem are the vectors $\vec{h_1},\ldots,\vec{h_r} \in \R^m$.
\end{theorem}

\begin{remark} \label{rem:geomean-SOCP}
The primal problem~\eqref{dualVmin} is equivalent to maximizing the geometric mean $\prod_{k=1}^r t_k^{\beta_k}$
under the same constraints, by monotonicity of the log function. Therefore, if $\vec{\beta}$ is rational
and the least common denominator of the $\beta_k$ is $q\leq 2^p$, we can reformulate Problem~\eqref{dualVmin}
as a SOCP by introducing less than $2^p$ additional norm constraints (cf. Section 6.2.3 of~\cite{NN94} or~\cite{LVBL98}).
 \end{remark}

The proof of Theorem~\ref{theo:SoptMulti} relies on a series of reformulations of the the $S_{\vec{\beta}}-$
optimal problem, relying on Lagrange duality techniques and Theorem~\ref{theo:SDPSol1}. We will prove this result in appendix.

Then, we will show that the optimality conditions of the present convex optimization problem
can be interpreted as geometrical conditions,
which yields a generalization of the theorem of Dette for $S-$optimality
to the case of multiresponse experiments~\cite{Dette93}. This geometrical characterization
relies on the following Elfving-type set:

$$\mathcal{D}_{\vec{\beta}}=\mathrm{convex\!-\!hull} \left\{\ \!\!\! \left(\begin{array}{c}
                                               \vec{\epsilon_1}^T A_{(1),\vec{x}} \\
						\vdots\\
						\vec{\epsilon_r}^T A_{(r),\vec{x}}\\
                                                 \end{array} \right) \!
,\ \vec{x} \in \mathcal{X},\ \vec{\epsilon_k}\in\R^l,\ \sum_{k=1}^r \beta_k \Vert \vec{\epsilon_k} \Vert^2 \leq 1 \right\}
\subset \R^{r \times m}.$$

\begin{theorem}[Geometrical characterization of multiresponse $S_{\vec{\beta}}-$optimality] \label{theo:DetteMulti}
The design  $\vec{w}$ is $S_{\vec{\beta}}-$optimal (and solution of Program~$(P_{\vec{\beta}})$) if and only if there exists
a vector $\vec{t} \in \R^r$ and vectors $\vec{\epsilon_{ik}} \in\R^{l}$ ($i \in [s], k \in [r]$),  such that
\begin{itemize}
 \item[(i)] $\forall i \in [s],\quad \sum_{k=1}^r \beta_k \Vert \vec{\epsilon_{ik}} \Vert^2 \leq 1$
 \item[(ii)] $\mathrm{Diag}(\vec{t}) C= \left(\begin{array}{c}
                                                  t_1 \vec{c_1}^T\\
						\vdots\\
						  t_r \vec{c_r}^T\\
                                                 \end{array} \right)=\sum_{i=1}^s w_i \left(\begin{array}{c}
                                                  \vec{\epsilon_{i1}}^T A_{(1),i}\\
						\vdots\\
						\vec{\epsilon_{ir}}^T A_{(r),i}\\
                                                 \end{array} \right)$
 \item[(iii)] $\mathrm{Diag}(\vec{t}) C$ lies on the boundary of $\mathcal{D}_{\vec{\beta}}$,
with a supporting hyperplane whose \emph{normal direction}
 is given by the matrix $H=[\vec{h_1},\ldots,\vec{h_r}]^T,$ with $\vec{h_k} \in \R^m$ . In other words,
$$D \in \mathcal{D}_{\vec{\beta}} \Longrightarrow \langle H,D \rangle \leq 1 $$ 
%(The latter inequality can also be written in vector form : if $H=[h_1,...,h_s]^T$ with $h_i \in \R^m$ and $\mathcal{D}_{\vec{\beta}} \ni D=[d_1,...,d_s]^T$ with $d_i \in \R^m$, then $\sum_i h_i^T d_i \leq 1$).
 \item[(iv)] $H$ satisfies the equalities $$t_k \vec{h_k}^T \vec{c_k}=\beta_k, \qquad \quad \forall k \in [r].$$
\end{itemize}

In this case, the optimal variables of Problems~$(D_{\vec{\beta}})$
and~$(P_{\vec{\beta}})$ are $\vec{t},\ \vec{v_{ik}}:=w_i \vec{\epsilon_{ik}}\ (\forall i \in[s],\ \forall k\in[r])$, and $(\vec{h_k})_{k \in [r]}$,
 so that the optimal $S_{\vec{\beta}}-$criterion is $-2 \sum_{k=1}^r \beta_k \log(t_k)$.
\end{theorem}

Theorem~\ref{theo:DetteMulti} is established in appendix.

\begin{remark}
%\textbf{1}\\
As in the case of single response experiments~\cite{Dette93}, the geometrical
characterization remains true when the regression range $~\mathcal{X}$ is infinite.
It can also be shown with semi-infinite programming techniques that the following convex semi-infinite program is valid for the general $S_{\vec{\beta}}-$optimal design
Problem:
\begin{align*}
\min _{\begin{array}{c}
        \scriptstyle{w_i\geq0, \sum_{i=1}^s w_i=1,}\\
    \scriptstyle{\vec{x} \in \mathcal{X}}
       \end{array}
} S_{\vec{\beta}}(\xi)=&\quad  2 \max_{\vec{h_1},\ldots,\vec{h_r}} \sum_{k=1}^r \beta_k \log \frac{\vec{c_k}^T \vec{h_k}}{\beta_k}\\
&\quad \forall \vec{x} \in \mathcal{X}, \quad \left\Vert \begin{array}{c} A_{(1),\vec{x}} \vec{h_1}/\sqrt{\beta_1} \\
					\vdots \\
					A_{(r),\vec{x}} \vec{h_r}/\sqrt{\beta_r}          
             \end{array} \right\Vert \leq 1. \nonumber
\end{align*}

\end{remark}

\begin{remark}
\label{rem:Dopt}
Dette showed in~\cite{Dette93} that $D-$optimality for the full parameter $\vec{\theta}$ is a particular case of $S-$optimality.
As a consequence, we can
formulate the $D-$optimal design problem as a convex optimization problem in the form of~$(P_{\vec{\beta}})$. To see this, 
Dette considered the virtual nested models, where the parameter of interest in the $k\th$ model is $\theta_k$,
and the observations only depend on the first $k$ parameters: $A_{(k),i}$ is the matrix $A_i$ restricted to its first $k$ columns,
so that $M_{(k)}(\vec{w})$ is the upper left $k \times k$ submatrix of $M(\vec{w})$, and $\vec{c_k}=[0,...,0,1]$ is a vector of length $k$.
Using the relation
$$\vec{c_k}^T M_{(k)}(\vec{w})^- \vec{c_k}=\big(M_{(k)}(\vec{w})^{-1}\big)_{kk}=\frac{\det M_{(k-1)} (\vec{w})}{\det M_{(k)} (\vec{w})},$$
it can be seen that $$S_{[1/m,...,1/m]}(\vec{w})=-\frac{1}{m} \log \det M(\vec{w}),$$
which is exactly the $D-$optimality criterion.\\
Theorem~\ref{theo:SoptMulti} can now be used to formulate the $D-$optimal design problem as:
\begin{align}
\max_{\vec{w}\geq\vec{0}}\ \frac{1}{m} \log \det M(\vec{w}) = & \quad 2 \max_{\vec{t}, \vec{v_{ik}}, \vec{w}}\ \log\Big( \big( \prod t_i \big)^{1/m} \Big) \nonumber\\
&\quad t_k \vec{c_k} = \sum_i A_{(k),i}^T \vec{v_{ik}},\qquad \quad  \quad \forall k \in [m], \label{Dopt-GP}\\
&\quad \left\Vert \begin{array}{c}
                   \vec{v_{i1}}\\
		   \vdots\\
		   \vec{v_{im}}
                  \end{array} \right\Vert \leq \sqrt{m}\  w_i\qquad \qquad \forall i \in [s], \nonumber\\
&\quad \sum_{i=1}^s w_i \leq 1. \nonumber
\end{align}
As pointed out in Remark~\ref{rem:geomean-SOCP}, this optimization problem can be reformulated
as a SOCP by introducing $n$ additional norm constraints, where $n$ is the smallest power of two that is greater
or equal to $m$.
\end{remark}

\section{Numerical Experiments} \label{sec:numexp}
In this section, we evaluate the benefits of the SOCP approach for the computation of optimal experimental designs.
We will see that the second order cone programs presented in this paper are very efficient when the number $r$ of
quantities to estimate is small (in particular for $\vec{c}-$optimality).% ($\vec{c}-$optimal design
%or $A-$optimal design for $K^T \vec{\theta}$, when $K$  has only a few columns).

We will compare the SOCP approach approach to 
the semidefinite programming/MAXDET approach~\cite{VBW98},
and to other classic algorithms from the experimental design literature.
We concentrate on Wynn--Fedorov-type exchange algorithms,
and Titterington-type multiplicative algorithms.
The former were discovered independently by Wynn~\cite{Wyn70} and Fedorov~\cite{Fed72}, and consist
in moving at each step toward the experiment with the largest directional derivative. Several versions and refinements
of this procedure were proposed; we will use the \emph{IncDec} procedure of Richtarik~\cite{Rich08}, which
specifies step lengths for which the precision $\delta$ is achieved in $O(1/ \delta)$ iterations.

On the other hand, multiplicative weight algorithms were introduced in 1976 by Titterington~\cite{Tit76}.
The monotonic behavior of any sequence generated by this algorithm was recently proved by Yu~\cite{Yu10a},
for a large class of design criteria, including $A-$ and $D-$optimality.
We will also consider a variant of the latter algorithm, for which Dette, Pepelyshev and Zhigljavsky~\cite{DPZ08}
%which relies on an acceleration parameter $\gamma$: they 
have established a convergence result in the case of $D-$optimality,
and conjectured the convergence for other criteria. These multiplicative algorithms use respectively a power parameter
$\lambda \in ]0,1]$ and an acceleration parameter $\gamma\in[0,1]$. We found that the values $\lambda=0.9$ and $\gamma=0.9$
gave the best results for $A-$optimality in our experiments, and so those values will be used throughout this section. For $D-$optimality,
we have used the acceleration parameter $\gamma=0.5$.

\begin{table}[ht!]
\begin{small}\begin{center}
\begin{tabular}{|c|ccccc|}
\hline
\multirow{2}{*}{$m$}	&	SOCP~\eqref{AP-SOCP} & SDP	&
	IncDec Exchange &	Accelerated Mult.  & Mult.Weight with	\\
		& [this paper]	&	\cite{VBW98}&	 \cite{Rich08}	& Weight ($\gamma=0.9$)~\cite{DPZ08} & Exponent $\lambda=0.9$~\cite{Yu10a}	\\\hline
$2$		& $  0.082$	& $  2.897$	& $ 10.039$	& $  3.026$	& $  2.979$	\\
$2^2$		& $  0.120$	& $  3.017$	& $ 99.510$	& $  9.598$	& $  9.240$	\\
$2^3$		& $  0.166$	& $  4.798$	& $ 13.112$	& $  5.883$	& $  6.040$	\\
$2^4$		& $  0.175$	& $  6.828$	& $ 24.431$	& $ 12.574$	& $ 12.204$	\\
$2^5$		& $  0.352$	& $ 15.820$	& $ 29.454$	& $ 11.258$	& $ 11.123$	\\
$2^6$		& $  0.816$	& $ 66.281$	& $ 54.379$	& $ 13.407$	& $ 13.419$	\\
$2^7$		& $  2.636$	& $338.669$	& $ 92.537$	& $ 37.935$	& $ 36.679$	\\
$2^8$		& $ 10.496$	& failed	& $202.509$	& $ 96.594$	& $ 99.751$	\\
$2^9$		& $ 44.689$	& failed	& $412.890$	& $585.619$	& $597.442$	\\
$2^{10}$	& $154.187$	& failed	& $498.616$	& $551.634$	& $539.130$	\\\hline
\end{tabular}
\end{center}\end{small}
\caption{CPU time (s) of the different algorithms, for  typical random instances of the $A-$optimal
design problem with $s=2^{10}$, $l=1$, $r=3$,
and different values of $m$.\label{tab:meffect}}
\end{table}
\begin{figure}[ht!]
\begin{center}
\includegraphics[width=\textwidth]{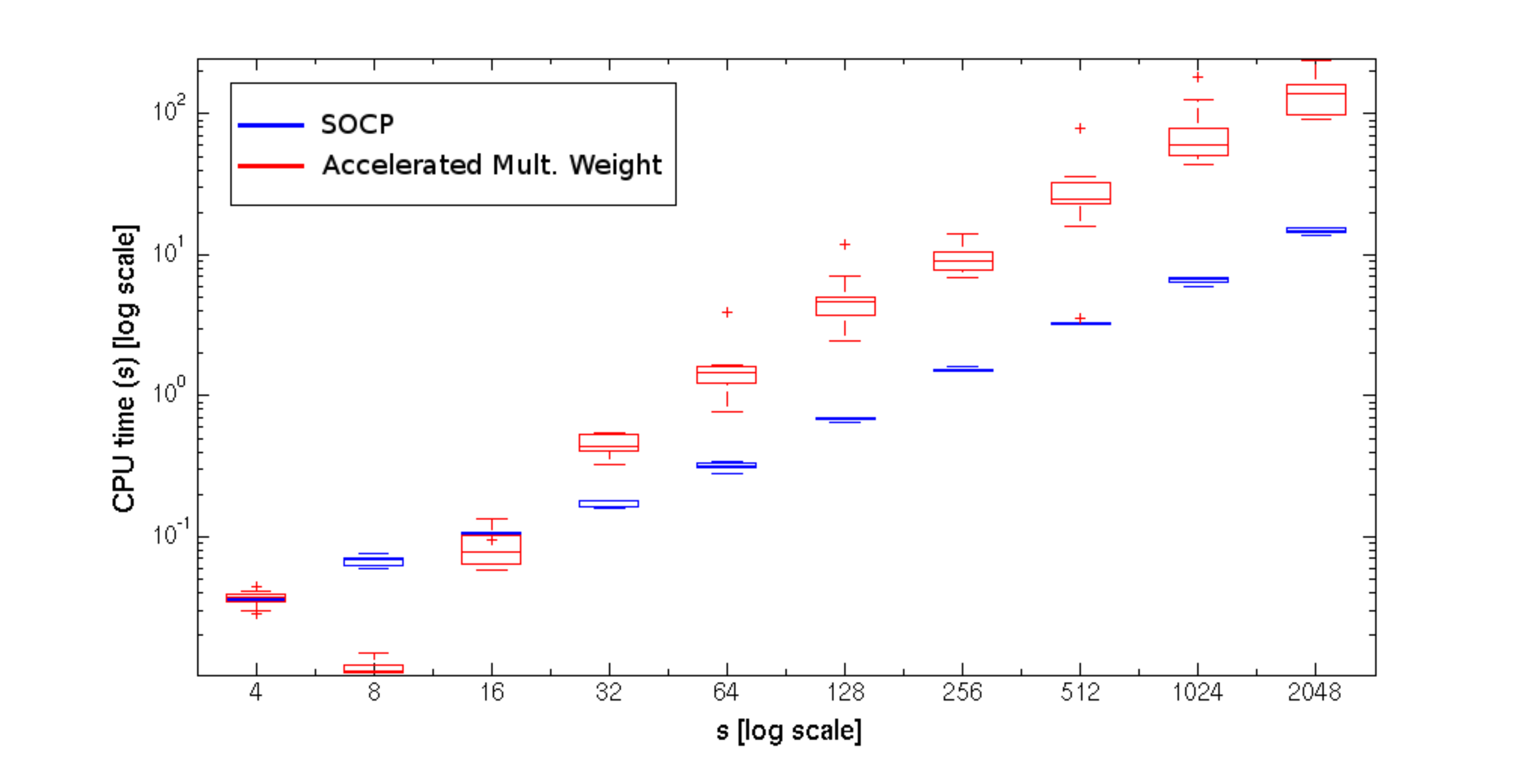} 
\end{center}
\caption{Comparison of two algorithms on random instances ($A-$optimality) with $m=120$,  $l=30$, $r=1$,
and varying $s$. The box plots represent the distribution of the computing times for $10$ random instances. \label{fig:seffect}}
\end{figure}

We will first consider random instances of optimal design problems, in order to evaluate to which extent
each parameter affects the computation time. Then, we will consider a simple polynomial regression
model, for which we shall see that the SOCP approach is well-suited when the number of support points is large.
%, and we will see that our approach is efficient even if the number of support points is large.
Finally we will present some results from a network application -- which is at the origin of this work-- where the sampling rates of a monitoring tool
should be optimized subject to multiple constraints.

\subsection{Random instances}

\begin{figure}[t]
\begin{minipage}{0.47\textwidth}
\begin{center}
\includegraphics[width=\textwidth]{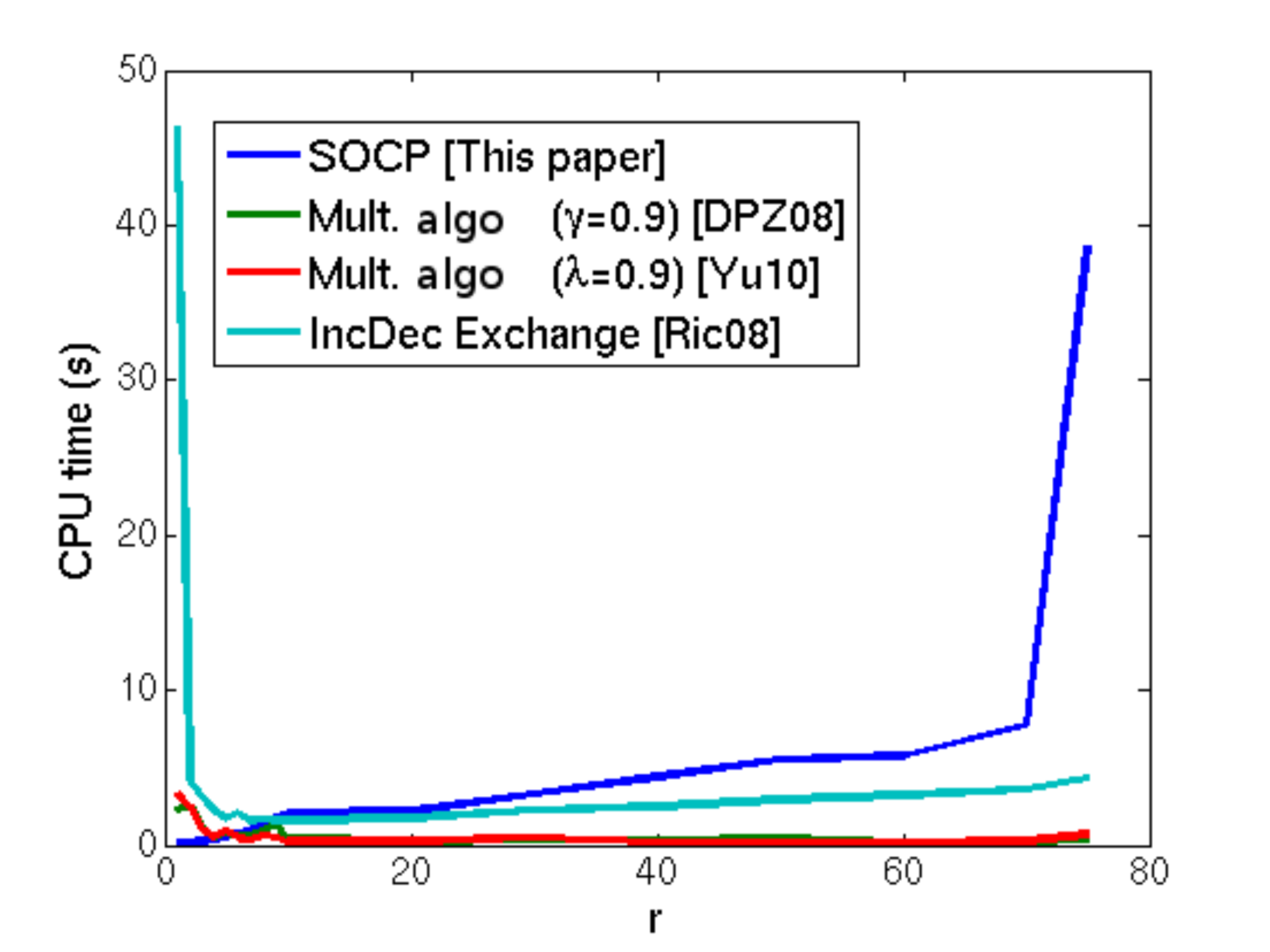} 
\end{center}
\caption{Comparison of four algorithms on typical random instances ($A-$optimality) with  $m=75$, $s=150$, $l=1$ and varying $r$.\label{fig:reffect}}
\end{minipage}
$\quad$
\begin{minipage}{0.47\textwidth}
\begin{center}
  \includegraphics[width=\textwidth]{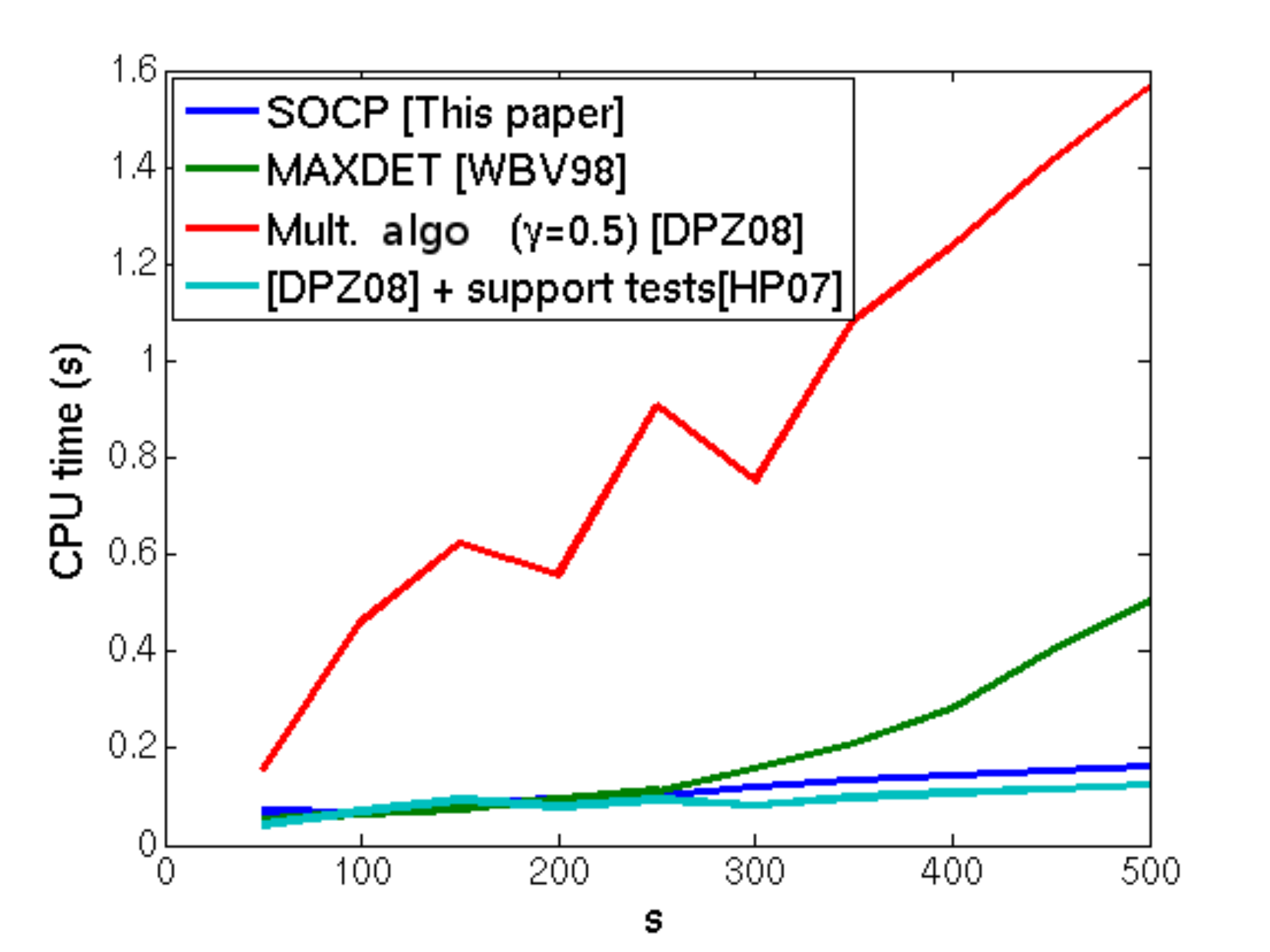} 
\end{center}
\caption{Comparison of four algorithms on typical random instances of the minimum covering ellipsoid ($D-$optimality for $\vec{\theta}$, $m=3$) and varying $s$. \label{fig:Dellipse}}
\end{minipage}
\end{figure}

In this section, we consider random instances of optimal experimental design problems, in which the entries of the $l\times m$ 
matrices $(A_i)_{i\in[s]}$
follow a normal distribution, as well as the entries of the $m \times r$ matrix $K$.
For every considered instance, we use SeDuMi to solve the SOCP~\eqref{AP-SOCP} and the $A-$optimality SDP~\cite{VBW98}; we
have implemented the other procedures in Matlab. In all our experiments, the stopping criterion is based on the general equivalence theorem of Kiefer~\cite{Kief74}:
the computation stops as soon as the ratio between the maximum of the gradient and the
value of the criterion is below 1.001 (as in~\cite{DPZ08}).

We start by evaluating the effect of $r$, which turns out to be the 
determining factor for the performance of the SOCP approach. To this end, we set $m=75$, $s=150$, $l=1$ (single-response experiments),
and we let $r$ vary between $1$ and $75$. The computing time of the different algorithms is plotted against $r$ in Figure~\ref{fig:reffect}.
We notice that the SOCP is the fastest for small values $(r\leq7)$, but performs badly
when $r$ is large, while the multiplicative weight algorithms are insensitive to the value of $r$.
For this reason, we will chose small values of $r$ in further experiments, since the SOCP approach might not be well adapted for large $r$.

We next study the effect of $s$ (the number of available experiments) for the case of $\vec{c}-$optimality ($r=1$). For these experiments, 
we set $m=120$,  $l=30$, and we take $s$ in the set $\{2^k, k=2,\ldots,11\}$. The performance (in terms of CPU time) of
the SOCP is compared to that of the accelerated multiplicative algorithm (for $\gamma=0.9$,~\cite{DPZ08})
on the log-log plot of Figure~\ref{fig:seffect}. The boxes represent the distribution of the CPU time, on $10$ randomly generated instances.
We see here that the SOCP approach is in average ten times faster as soon as $s\geq32$.

To evaluate the effect of $m$, we set  $s=2^{10}$, $l=1$, $r=3$, and choose $m$ in the set $\{2^k, k=1,\ldots,10\}$ (Note
that since $l=1$ and $K$ is randomly generated, we must have $s\geq m$ for the instance to be feasible). The results
of each algorithm are displayed in Table~\ref{tab:meffect}. It is striking that the SOCP approach is the best one, while the SDP
is the worst when $m$ becomes large, which demonstrates the importance of the rank reduction discussed in Section~\ref{sec:lagrange}.
For $m\leq2^9$, the SOCP is 10 times faster than all other algorithms. In the last row of the table however, this ratio is lower.
This might be because $s=m=2^{10}$ in this case, such that all experiments are support points of the optimal design, and 
classic algorithms certainly take advantage of this situation (while it does not make a difference for interior point codes).

Pronzato~\cite{Pro03} has shown that we can improve the multiplicative algorithms thanks to
a simple test which allows to remove \emph{on the fly} experiments which do not belong to the $D-$optimal design (i.e.\ with a zero weight),
and which was refined by Harman and Pronzato~\cite{HP07}. This can considerably improve the performance of
the multiplicative algorithms when there are a lot of points with a zero weight. As in~\cite{HP07}, we have studied random instances
of the minimum covering ellipse, but in $\R^3$: $m=3$, $K=\vec{I}_3$, and we draw $s$ independent
random regression vectors ($l=1$) from a normal distribution $\vec{a_i}\sim\mathcal{N}(0,\vec{I}_3)$, with
$s$ increasing from $50$ to $500$. The $D-$optimal design problem is equivalent to finding the minimum volume ellipsoid which contains
the $s$ vectors $\vec{a_i}$, and the $D-$optimal design is supported by points lying on the boundary of this minimal ellipsoid.
In accordance with intuition, the number of support points of the $D-$optimal design is small, and therefore the test of
Pronzato and Harman improves dramatically the computing time~(cf.\ Figure~\ref{fig:Dellipse}). Note however that the
SOCP for $D-$optimality~\eqref{Dopt-GP} remains competitive with the latter approach.

\subsection{Polynomial Regression}

\begin{figure}[ht]
\begin{center}
  \includegraphics[width=\textwidth]{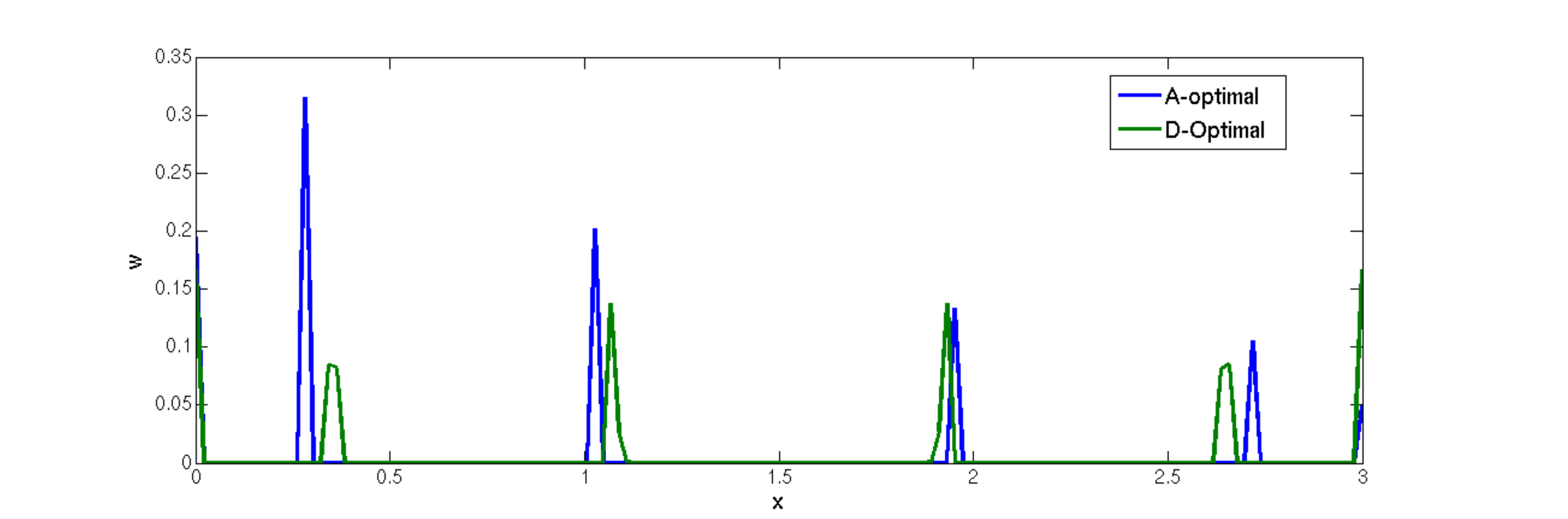} 
\end{center}
\caption{A- and D-optimal designs for the polynomial regression model of degree $5$ on $\mathcal{X}=[0,3]$. \label{fig:AoptDopt}}
\end{figure}

\begin{figure}[ht]
\begin{minipage}{0.48\textwidth}
\begin{center}
  \includegraphics[width=\textwidth]{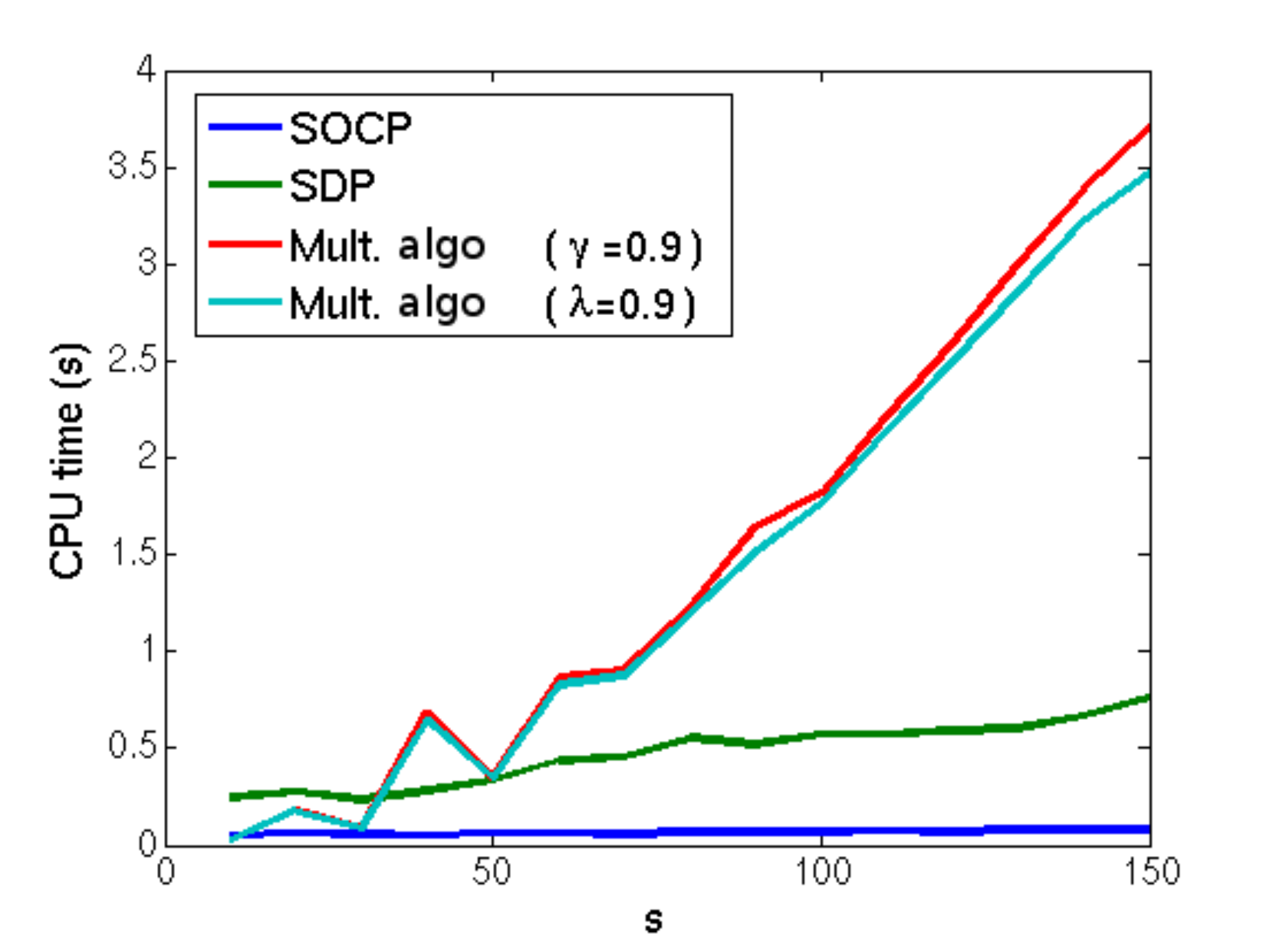} 
\end{center}
\caption{A-optimal design for the polynomial regression model: evolution of the computation time with the number of points
for the discretization of $[0,3]$.
\label{fig:Aopt}}
\end{minipage}
$\quad$
\begin{minipage}{0.42\textwidth}
\vspace{6mm}
\begin{center}
\includegraphics[width=\textwidth]{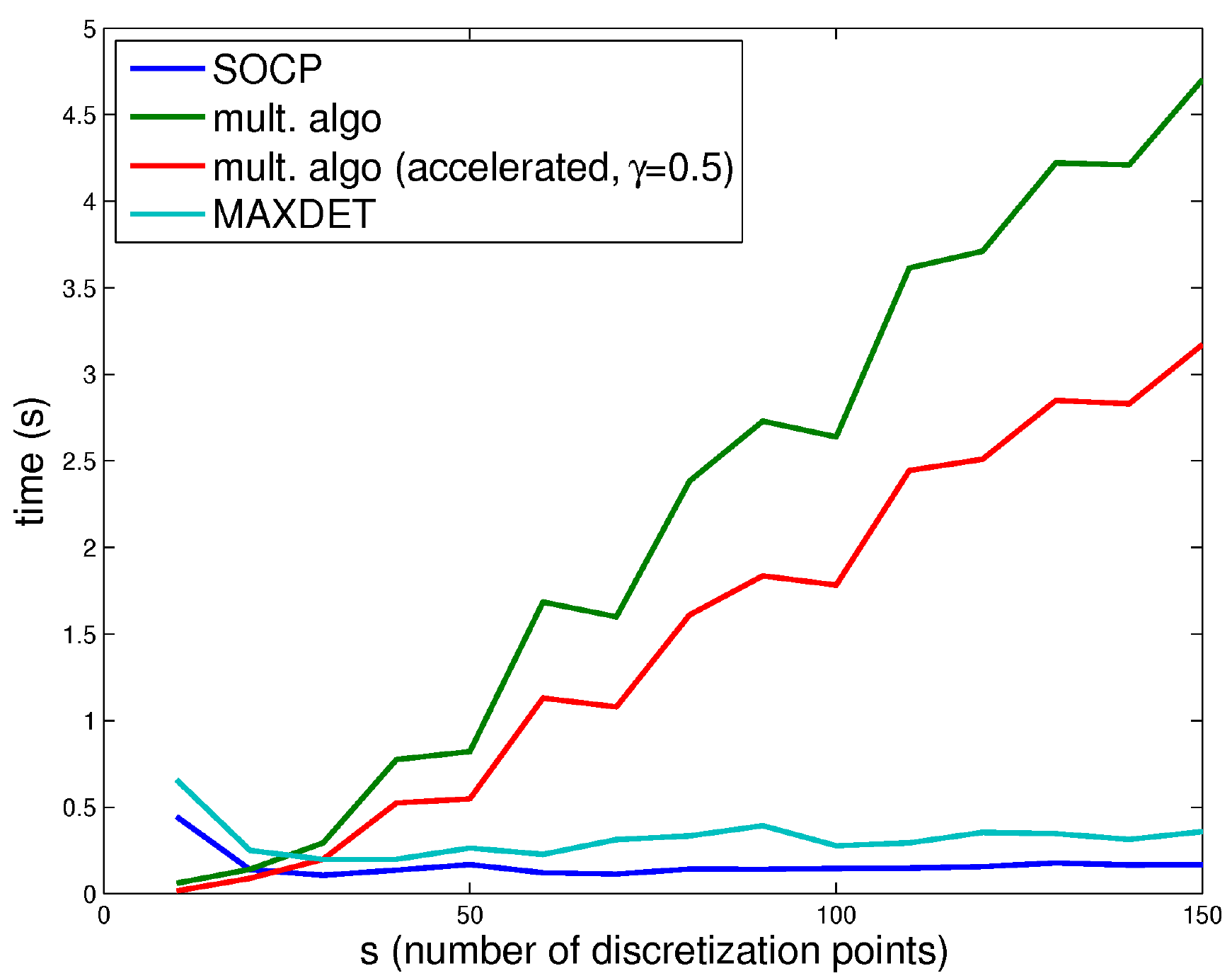} 
\end{center}
\caption{D-optimal design for the polynomial regression model: evolution of the computation time with the number of points
for the discretization of $[0, 3]$.
\label{fig:Dopt}}
\end{minipage}
\end{figure}

We have computed the $A-$ and $D$-optimal designs (for the full parameter $\vec{\theta}$),
for a polynomial regression model of degree $5$:
$$A(\vec{\vec{x}})=[1,\vec{x},\vec{x}^2,\vec{x}^3,\vec{x}^4,\vec{x}^5]$$
on the regression region $\mathcal{X}=[0,3]$. The optimal designs are represented on Figure~\ref{fig:AoptDopt}. In this problem,
we have $r=m=6$, which is \emph{small}. Therefore, we can hope that the SOCP approach will perform well. The computation
times are plotted on Figures~\ref{fig:Aopt} and~\ref{fig:Dopt}, as a function of the number of points considered for the discretization of
the regression interval $\mathcal{X}=[0,3]$.
For the $A-$optimal design, the experimental setting was the same that the one of previous section. For the $D-$optimal design problem,
we solved the geometric program~\eqref{D-SOCP} with SeDuMi. We have also implemented the classic multiplicative algorithm,
the accelerated algorithm with $\gamma=0.5$, and the MAXDET program~\cite{VBW98}. Contrarily to the multiplicative algorithms,
the interior point algorithms (SOCP and MAXDET) seem to be insensitive to the size of the discretization grid. For these instances,
the SOCP is roughly two times faster than the MAXDET program. Also note that the effect of the acceleration parameter $\gamma$ is clearly
visible (red curve vs.\ green curve).
We point out that for these polynomial regression problems,
the tests of Pronzato and Harman~\cite{Pro03,HP07} to remove points that do not belong to the
support of the $D-$optimal design did not yield any improvement.

\subsection{Optimal Sampling in IP networks}
\label{sec:numexpNetflow}

We finally show some results for an application to the optimal monitoring of large IP networks.
Assume that an Internet provider wants to estimate the traffic matrix of his network, that is, the
volume of traffic between each pair of origin and destination during a given
time period. To this end, he disposes of a monitoring tool, which can be activated at different sampling rates
in different location of the network, and is able to find the destination of the sampled packets.
For networking issues, the intensive use of this tool is not suitable, because it creates an overload
both in terms of CPU utilization of the router and bandwidth consumption. The sampling rates should therefore
be tuned cautiously on each interface, in such a way that the number of sampled packets remains under a target threshold.

This situation can be represented by an optimal design model with multiresponse experiments: the set of available experiments
$\mathcal{X}$ coincides with the interfaces of the network where the monitoring-tool can be activated: when the software is installed
on a given interface, we obtain an estimation of the sum of the flows that traverse this interface, and that have destination $D$, for
every destination $D$ reachable from this interface. In~\cite{SagnolGB10ITC}, two coauthors and I have shown that if the sampling rates
are small, then the Fisher information matrix of the sampling design has the standard form~\eqref{infoMat}
(after an appropriate normalization of the observation matrices relying on a prior estimate of the unknown OD traffic matrix).
The optimal monitoring problem can thus be formulated as an optimal experimental design problem with multiple resource constraints.

We first study some $\vec{c}-$optimal sampling problems with the simple constraint $\sum_{i=1}^s w_i =1$,
so that we can compare the present SOCP approach to classic algorithms.
Table~\ref{tab:NetProb} summarizes the results (in terms of CPU time)
for several problems: each instance is defined by a network and the type of interfaces considered.
We used the
topology of three networks: Abilene, which consists in $11$ nodes, $m=121$ OD pairs and  $50$ links; the Opentransit
backbone of France Telecom, with $116$ nodes, $m=13456$ OD pairs and $436$ links; and a clustered version of the latter
network, thus reduced to $31$ nodes, $m=961$ OD pairs and $133$ links. The natural problem is to activate the monitoring tool
independently on each link (interface=``links''). However, we also considered the academic problem of imposing the same sampling rates on all
incoming links of each router, which is equivalent to consider each router as a \emph{big interface} (interface=``Nodes''). 
For all these instances the vector $\vec{c}$ was drawn from a normal distribution.
The threshold for the stopping criterion was lowered to $1.01$ for this network application, since this value suffices to obtain good designs in practice.

\begin{table}[h]
\begin{small}\begin{center}
\begin{tabular}{@{}c|cccccc@{}}
\multirow{2}{*}{Network}	& Abilene &	Abilene	& OTClusters &	OTClusters & Opentransit & Opentransit\\
		& ($m=121$)	&($m=121$)	&($m=961$)	&($m=961$)&($m=13456$) & ($m=13456$)	\\\hline
\multirow{2}{*}{Interfaces}	&  Nodes  & Links & Nodes  & Links & Nodes  & Links  	\\
                &   ($s=11$)   &  ($s=50$)    &  ($s=31$)     &   ($s=133$)     &   ($s=116$)    & ($s=436$) \\ \hline
SOCP	            & $  0.021$	& $  0.036$	& $  0.078$	& $  0.094$	& $5.52$ & $33.03$ \\
SDP		    & $  1.095$	& $  1.178$	& $692.37$	& $734.25$	& failed     & failed\\
IncDec Exchange     & $0.518$   & $0.823$    & $4.57$     & $19.69$ & failed & failed\\
Mult. algo ($\gamma=0.9$)&$0.009$ & $  0.043$	& $  0.018$	& $  1.893$	& failed     & failed\\
Mult. algo ($\lambda=0.9$)&$0.008$& $  0.038$	& $  0.018$	& $  1.468$	& failed     & failed\\
\end{tabular}
\end{center}\end{small}
\caption{CPU time (s) for different instances of $\vec{c}-$optimal design arising from an optimal monitoring
problem in IP networks (with the standard constraint $\sum_i w_i=1$) \label{tab:NetProb}}
\end{table}

We can see in the table that the multiplicative algorithms perform better than the SOCP approach on the
instances where $s$ is small (1st and 3rd columns in Table~\ref{tab:NetProb}). On the other instances
however, the SOCP performs well, and it is the only method which returned a solution for the
Opentransit network. The SDP and the multiplicative methods failed because of memory issues
(in the multiplicative algorithm, a full rank update of the $13456 \times 13456$ information
matrix should be carried out at each time step). The IncDec Exchange algorithm did not crash,
but it had not converged after 2 hours of computation.

We next turn to the case of general constraints of the form $R\vec{w}\leq \vec{b}$. Since we do not know any other algorithm which can handle optimal design problems with multiple resource constraints, we compare
the SOCP and the semi-definite programming approaches only. Table~\ref{tab:consProb} summarize the results (in terms of CPU time)
for several problems, specified as previously by the network and the type of interfaces considered,
and also by the type of the constraint matrix $R$. 
In the optimal sampling problem, the matrix $R$ usually depends on the volume of traffic observed at each router (cf.~\cite{SagnolGB10ITC}). We
simulated this data from a uniform distribution, a lognormal distribution, or we used real traffic loads.
To see the effect of the number of
constraints, we also generated arbitrary constraints matrices of different sizes.

In comparison to the SDP, the computation time can be reduced by a factor in the order of $1000$ on
the instances from the clustered network.
Moreover, the SOCP approach is able to handle huge instances arising from the Opentransit network (in which $m=13456$).
\begin{table}[ht]
\begin{small}\begin{center}
\begin{tabular}{@{}c|ccccc@{}}
\multirow{2}{*}{Network}& Abilene	&	Abilene		&	 Abilene 	&	Abilene	 	&	 Abilene\\
		& ($m=121$)		&	($m=121$)	&	($m=121$)	&	($m=121$)	&	($m=121$)	\\\hline
Interfaces	& Links ($s=50$)	& Links ($s=50$)	&	Links ($s=50$)	&	Nodes ($s=11$)	& Nodes ($s=11$)	\\\hline
\multirow{2}{*}{Constraints}		& R: $11 \times 50$ 	& R: $11 \times 50$	& R: $11 \times 50$	& R: $4 \times 11$	&	R: $10 \times 11$\\
		& (uniform traffic)	& (lognormal traffic)	& (real traffic)	& (arbitrary)		&	(arbitrary) \\
SOCP		& $  0.043$		& $  0.056$		& $0.061$		& $  0.051$		& $  0.053$	\\
SDP		& $  0.714$		& $  0.842$		& $0.944$		& $  0.827$		& $  0.876$	\vspace{1cm}\\
\multirow{2}{*}{Network}& OTClusters	& OTClusters 		& OTClusters		& Opentransit		& Opentransit	\\
		& ($m=961$)		&	($m=961$)	&	($m=961$)	&	($m=13456$)	&	($m=13456$)	\\ \hline
Interfaces	&	Nodes ($s=31$)	& Links ($s=133$)	&	Links ($s=133$)	&	Links ($s=436$)	&	Links ($s=436$)\\ \hline
\multirow{2}{*}{Constraints}		& R: $4 \times 31$	& R: $31 \times 133$	& R: $130 \times 133$	& R: $12 \times 436$	&	R: $116 \times 436$ \\
		& (arbitrary)		&(uniform traffic)	& (arbitrary)		& (arbitrary)		&	(real traffic) \\
SOCP		& $ 0.141$		& $0.462$		& $1.135$ 		& $23.32$		& $187.59$	\\
SDP		& $350.63$		& $451.69$		& $430.71$		& failed		& failed
\end{tabular}
\end{center}\end{small}
\caption{Computation time (s) for different instances of $\vec{c}-$optimal design arising from an optimal monitoring
problem in IP networks (with  multiple constraints $R\vec{w}\leq \vec{b}$).\label{tab:consProb}}
\end{table}

%TODO Conclusion ?
\begin{section}{Acknowledgment}
The author is immensely grateful to St\'ephane Gaubert, whose  comments,
advice and support were essential. He also wants to thank Mustapha Bouhtou for the
stimulating discussions which are at the origin of this work.
He also expresses his gratitude to two anonymous referees for their constructive comments, and more particularly
to a referee who gave precious suggestions for the presentation of these experimental results.
\end{section}

%------------------------------------------
%      APPENDIX
%------------------------------------------

% 

\clearpage

\renewcommand \appendixname{}
\appendix
\noindent{\textbf {\Large Appendix}}

\section{Proofs of Theorems~\ref{theo:SoptMulti} and~\ref{theo:DetteMulti}}
\label{sec:proofs}

We start with the following lemma, where we show that the $S_{\vec{\beta}}-$optimal design problem can be formulated as a convex
optimization problem with SDP constraints:

\begin{lemma} \label{lem:P/Dbeta-SDP}
The optimal variable $\vec{w^*}$ of the following convex optimization problem also minimizes the $S_{\vec{\beta}}-$criterion. The value of
this program coincides with the value of its dual, which we give below:
\begin{align}
\min_{\vec{w}\geq\vec{0}, \sum_i w_i=1} S_{\vec{\beta}}(\vec{w})= & \min_{\vec{\tau} \in \R^r,\ \vec{w}\geq\vec{0}}
    - \sum_{k=1}^r \beta_k \log \tau_k \tag{$P_{\vec{\beta}}-SDP$} \label{primalLogSDP}\\
&\quad M_{(k)}(\vec{w}) \succeq \tau_k \vec{c_k} \vec{c_k}^T,\qquad \forall k \in [r], \nonumber\\
&\quad \sum_{i=1}^s w_i =1.\nonumber\\
~ \nonumber \\
=& \max_{Z_1,\ldots, Z_r\succeq0} \quad \sum_{k=1}^r \beta_k \log \frac{\vec{c_k}^T Z_k \vec{c_k}}{\beta_k} 
	 \tag{$D_{\vec{\beta}}-SDP$} \label{dualLogSDP}\\
&\quad \sum_{k=1}^r \operatorname{trace} (A_{(k),i}\: Z_k\: A_{(k),i}^T) \leq 1,\qquad \forall i \in [s]. \nonumber
\end{align}

\end{lemma}

\begin{proof}
As in the proof of Theorem~\ref{theo:SDP}, we reexpress the variance of the $k\th$ quantity of interest
$\vec{c_k}^T M_{(k)}(\vec{w})^- \vec{c_k}$ with the help of a generalized Schur complement (for an arbitrary design $\vec{w}$):
\begin{equation*}
\begin{array}{ccc}
\big(\vec{c_k}^T M_{(k)}(\vec{w})^- \vec{c_k}\big)^{-1}=&\max \quad \tau_k			&=\quad \max \quad \tau_k\\
				&\ \left( \begin{array}{c|c}		
				               M_{(k)}(\vec{w}) & \vec{c_k}\\
						\hline
						\vec{c_k}^{{}_T}   & 1/\tau_k
				              \end{array}\right) \succeq 0.	&\qquad \qquad \qquad M_{(k)}(\vec{w}) \succeq \tau_k \vec{c_k} \vec{c_k}^T.
\end{array}
\end{equation*} 
As pointed out in the proof of Theorem~\ref{theo:SDP}, we can always assume that $\tau_i>0$ if we exclude the trivial case $\vec{c_k}=\vec{0}$,
 so that the latter expression is well defined. Now, by monotonicity of the log function, we can write:
\begin{align*}
\min_{\vec{w}\geq\vec{0}, \sum_i w_i=1} S_{\vec{\beta}}(\vec{w})=& - \max_{\vec{w}\geq\vec{0}, \sum_i w_i=1}\quad
 \sum_{k=1}^r \beta_k \log \big(\vec{c_k}^T M_{(k)}(\vec{w})^- \vec{c_k}\big)^{-1}\\
=&- \max_{\vec{w}\geq\vec{0}, \sum_i w_i=1,\ \vec{\tau}\in \R^r} \qquad \sum_{k=1}^r \beta_k \log \tau_k\\
&\qquad \quad \quad M_{(k)}(\vec{w}) \succeq \tau_k \vec{c_k} \vec{c_k}^T,\qquad \forall k \in [r], \nonumber
\end{align*}
which is exactly Problem~$(P_{\vec{\beta}}-SDP)$. It is clear that Problem~$(D_{\vec{\beta}}-SDP)$ is convex and strictly feasible,
so that the
Slater condition is fulfilled, and strong duality holds. It remains to show that Problem~$(D_{\vec{\beta}}-SDP)$ is indeed the dual
of~$(P_{\vec{\beta}}-SDP)$. To this end, let us form the Lagrangian of Problem~$(P_{\vec{\beta}}-SDP)$:
$$\mathcal{L}\big((\vec{\tau},\vec{w}),(Z,\lambda)\big)=-\sum_{k=1}^r \beta_k \log \tau_k + \sum_{k=1}^r \langle Z_k , \tau_k \vec{c_k} \vec{c_k}^T - M_{(k)}(\vec{w}) \rangle
+\lambda (\sum_{i=1}^s w_i-1).$$
The Lagrange dual function is given by
\begin{align*}
g(Z,\lambda)&:= \min_{\vec{\tau}>\vec{0},\ \vec{w}\geq\vec{0}}\ \mathcal{L}\big((\vec{\tau},\vec{w}),(Z,\lambda)\big)\\
&=-\lambda + \sum_k \min_{\tau_k>0} (\tau_k \vec{c_k}^T Z_k \vec{c_k} - \beta_k \log \tau_k)
	 + \sum_i \min_{\vec{w} \geq \vec{0}}\ w_i(\lambda - \sum_k \langle A_{(k),i}^T\ A_{(k),i}, Z_k \rangle).\\
&=\left\lbrace\begin{array}{cl}
-\lambda+\sum_k \beta_k (1- \log \frac{\beta_k}{\vec{c_k}^T Z_k \vec{c_k}}) & \mathrm{if}
\left\{
\begin{array}{l}
\quad \forall i \in [s],\ \sum_{k=1}^r
		     \langle A_{(k),i}^T\ A_{(k),i}, Z_k \rangle \leq\lambda \\
\quad \forall k \in[r],\ \vec{c_k}^TZ_k\vec{c_k}>0
\end{array} \right.\\
-\infty & \mathrm{otherwise.}
  \end{array}\right.
\end{align*}
Note that in the above expression, the minimum over $\tau_k$ is attained for $\tau_k=\frac{\beta_k}{\vec{c_k}^T Z_k \vec{c_k}}$,
and this equation must be satisfied by the optimal variables $\tau_k^*$ and $Z_k^*$. Since we observed that strong duality holds,
the value of the dual optimization problem must be equal to the value of the primal,
and so the optimal variables (denoted with stars in superscript) satisfy:
$$- \sum_{k=1}^r \beta_k \log \tau_k^* =  -\lambda^* +\sum_k \beta_k (1- \log \frac{\beta_k}{\vec{c_k}^T Z_k^* \vec{c_k}} )
\qquad \Longrightarrow \lambda^*=\sum_{k=1}^r \beta_k=1.$$
We can now make the dual problem explicit:
\begin{align*}
\max_{Z,\lambda} g(Z,\lambda) = \max_{Z_1,\ldots,Z_r \succeq 0} g(Z,1) =& \max_{Z_1,\ldots Z_s\succeq0} \quad 
\sum_{k=1}^r \beta_k \log \frac{\vec{c_k}^T Z_k \vec{c_k}}{\beta_k} \\
&\quad \sum_{k=1}^r \operatorname{trace} (A_{(k),i}\ Z_k\ A_{(k),i}^T) \leq 1,\qquad \forall i \in [s]. \nonumber
\end{align*}
This completes the proof of the lemma.
\end{proof}

Now, we show that there is a solution of Problem~$(D_{\vec{\beta}}-SDP)$ for which every $Z_k$ has rank one,
 thanks to the theoretical result presented in Section~\ref{sec:lagrange}.

\begin{proof}[Proof of Theorem~\ref{theo:SoptMulti}]
We first write the program~$(D_{\vec{\beta}}-SDP)$ in the form of a separable optimization problem, by introducing some vectors
$\vec{\alpha_i}$ ($i \in [s]$) of size $r$, satisfying $\sum_{k=1}^r \alpha_{ik} \leq 1$:
\begin{align*}
\min_{\vec{w}\geq\vec{0}, \sum_i w_i=1} S_{\vec{\beta}}(\vec{w}) =& \max_{\vec{\alpha_1},\ldots \vec{\alpha_s}\in \R^r}
\qquad  \qquad \left( \sum_{k=1}^r f_k(\alpha_{1k},\ldots,\alpha_{sk}) \right)\\
&\forall i \in [s],\ \sum_{k=1}^r \alpha_{ik} \leq 1, \nonumber
\end{align*}

where we have set 
\begin{align*}
\forall k \in [r],\qquad f_k(\vec{y})&=\max_{Z_k\succeq 0}\quad  \beta_k \log \frac{\vec{c_k}^T Z_k \vec{c_k}}{\beta_k} \\
&\qquad \operatorname{trace} (A_{(k),i}\ Z_k\ A_{(k),i}^T) \leq y_i,\ \quad \forall i \in [s].
\end{align*}

By use of Theorem~\eqref{theo:SDPSol1} (and monotonicity of the log function),
the minimization problem over $Z_k$ in $f_k(\cdot)$ has a rank-one solution ($Z_k=\vec{h_k} \vec{h_k}^T$), and
we obtain:
\begin{align*}
f_k(\alpha_{1k},\ldots,\alpha_{sk})&=\max_{\vec{h_k}\in \R^m}\quad  \beta_k \log \frac{(\vec{c_k}^T \vec{h_k})^2}{\beta_k} \\
&\qquad \Vert A_{(k),i}\ \vec{h_k} \Vert \leq \sqrt{\alpha_{ik}},\ \quad \forall i \in [s].
\end{align*}

Now, we use the associativity of the maximum to reformulate the $S_{\vec{\beta}}-$optimum design problem:
\begin{align*}
\min_{\vec{w}\geq\vec{0}, \sum_i w_i=1} S_{\vec{\beta}}(\vec{w}) =& \quad  \max_{\vec{h_1},\ldots,\vec{h_s}}
\sum_{k=1}^r \beta_k \log \frac{(\vec{c_k}^T \vec{h_k})^2}{\beta_k}\\
&\quad \sum_{k=1}^r \Vert  A_{(k),i}\ \vec{h_k} \Vert^2 \leq 1,\quad \forall i\in[s]. \nonumber
\end{align*}

Finally, we make the change of variable $\vec{h_k}'=\vec{h_k} \sqrt{\beta_k}$ in order to obtain the desired optimization problem, that is~$(D_{\vec{\beta}})$.
It remains to show that Problem~$(P_{\vec{\beta}})$ is the dual of~$(D_{\vec{\beta}})$. The convex problem~$(P_{\vec{\beta}})$
is strictly feasible, so that Slater condition is fulfilled, and strong duality holds.\\
We will now dualize Problem~$(P_{\vec{\beta}})$. This part of the proof is very similar to the dualization of Problem~$(D_{\vec{\beta}}-SDP)$ of
the previous
lemma. We include it here, though, for the reader's convenience. In the sequel, we denote by $\vec{v_i}$ the concatenation of the vectors
$\vec{v_{ik}}$:~$\vec{v_i}=[\vec{v_{i1}}^T,\ldots,\vec{v_{ir}}^T]^T \in \R^{rl},$ and by $\vec{\tilde{\beta}}$ the vector
containing $\beta_k$
entries arranged in blocks of
length $l$: $\vec{\tilde{\beta}}=[\beta_1,\ldots,\beta_1,\ldots,\ldots,\beta_r,\ldots,\beta_r]^T \in \R^{rl}.$
We also use the symbol $\odot$ to denote the Hadamard
product of vectors (elementwise product). With this notation, we can write:
$$\left( \begin{array}{c}
                   \sqrt{\beta_1} \vec{v_{i1}}\\
		   \vdots\\
		   \sqrt{\beta_r} \vec{v_{ir}}
                  \end{array} \right)=\vec{\tilde{\beta}}^{1/2} \odot \vec{v_i}.$$
We denote by $V$ the family of vectors $(\vec{v_{ik}})_{i \in [s],\ k \in [r]}$ and by $H$ the family of vectors $(\vec{h_k})_{i\in[s]}$.
Now, let us form the Lagrangian
\begin{eqnarray}
\mathcal{L}\big( (\vec{t},V,\vec{w}),(H,\vec{\mu},\lambda) \big)= \sum_{k=1}^r -\beta_k \log t_k + \sum_{k=1}^r \vec{h_k}^T
(t_k \vec{c_k} - \sum_{i=1}^s A_{(k),i}^T \vec{v_{ik}})\\
  +\sum_{i=1}^s \mu_i (\Vert \vec{\tilde{\beta}}^{1/2} \odot \vec{v_{i}} \Vert -w_i) + \lambda (\sum_{i=1}^s w_i-1) \nonumber
\end{eqnarray}
 The Lagrange dual function is given by
\begin{align*}
 g(H,\vec{\mu},\lambda)&:= \min_{\vec{t},V,\ w} \mathcal{L}\big( (\vec{t},V,\vec{w}),(H,\vec{\mu},\lambda) \big) \\
&=-\lambda + \sum_{k=1}^r  \min_{t_k} (t_k \vec{h_k}^T \vec{c_k} - \beta_k \log t_k )+ \sum_{i=1}^s \min_{w_i} w_i (\lambda-\mu_i) \\
&\qquad +   \sum_{i=1}^s \min_{\vec{v_{i}}} (\mu_i \Vert \vec{\tilde{\beta}}^{1/2} \odot \vec{v_{i}} \Vert -\vec{z_i}^T \vec{v_{i}}),
\end{align*}
where we have defined the vectors $\vec{z_i}^T := [\vec{h_1}^T A_{(1),i}^T ,...,\vec{h_r}^TA_{(r),i}^T] \in \R^{rl}$. In the latter equation,
the minimum over $t_k$ is finite
if and only if $\vec{c_k}^T \vec{h_k}>0$, and is attained
for $t_k=\frac{\beta_k}{\vec{c_k}^T \vec{h_k}}$; the expression in $w_i$ is bounded from below (by $0$)
 if and only if $\mu_i = \lambda$. The
reader can also verify that the minimization with respect to $\vec{v_{i}}$ is unbounded whenever
$\Vert \vec{\tilde{\beta}}^{-1/2} \odot \vec{z_i}\Vert > \mu_i$, and
takes the value $0$ otherwise. The Cauchy Schwarz inequality between the vectors  $\vec{\tilde{\beta}}^{-1/2} \odot \vec{z_i}$
and $\vec{\tilde{\beta}}^{1/2} \odot \vec{v_i}$ shows indeed
that the minimum is attained for a vector such that $\vec{v_i}$ is proportional to $\vec{\tilde{\beta}}^{-1} \odot \vec{z_i}$
if $\Vert \vec{\tilde{\beta}}^{-1/2} \odot \vec{z_i}\Vert = \mu_i$, and for $\vec{v_i}=\vec{0}$
if $\Vert \vec{\tilde{\beta}}^{-1/2} \odot \vec{z_i}\Vert < \mu_i$. To summarize,
\begin{equation*}
 g(H,\vec{\mu},\lambda)=\left\lbrace \begin{array}{cl}
                    -\lambda+\sum_{k=1}^r \beta_k (1-\log\frac{\beta_k}{\vec{c_k}^T \vec{h_k}})	& \textrm{if } \forall i\in[s],\
													\mu_i=\lambda \textrm{ and }
													\Vert \vec{\tilde{\beta}}^{-1/2} \odot \vec{z_i}\Vert \leq \mu_i;\\
		    -\infty			& \textrm{otherwise.}
                   \end{array} \right.
\end{equation*}

Now, since the primal and the dual share the same optimal value (we observed that strong duality holds), it follows that the optimal variables (denoted with stars in superscript) satisfy
$$g(H^*,\vec{\mu^*},\lambda^*)=-\lambda^*+\sum_{k=1}^r \beta_k (1-\log\frac{\beta_k}{\vec{c_k}^T \vec{h_k}^*})=
 \sum_{k=1}^r -\beta_k \log t_k^*.$$
Combining this equality with the stationarity equations $t_k^*=\frac{\beta_k}{\vec{c_k}^T \vec{h_k}^*}$ and
$\mu_i^*=\lambda^*$, we obtain:
$$\lambda^*=\mu_i^*=\sum_{k=1}^r \beta_k =1, \qquad \qquad \forall i\in[s].$$
We can now make the dual of~$(P_{\vec{\beta}})$ explicit:
\begin{align}
\min _{\vec{w}\geq\vec{0}, \sum w_i=1} S_{\vec{\beta}}(\vec{w}) = & \quad
       2 \max_{H} \sum_{k=1}^r \beta_k \log \frac{\vec{c_k}^T \vec{h_k}}{\beta_k} \nonumber\\
&\quad \vec{z_i} = \left( \begin{array}{c} A_{(1),i} \vec{h_1} \\
					\vdots \\
					A_{(r),i} \vec{h_r}          
             \end{array} \right) \nonumber,\quad \forall i \in [s],\\
&\quad \Vert \vec{\tilde{\beta}}^{-1/2} \odot \vec{z_i} \Vert \leq 1,\qquad \: \forall i \in [s].\nonumber
\end{align}
This program is the same as~$(D_{\vec{\beta}})$, and it completes the proof of Theorem~\ref{theo:SoptMulti}.
\end{proof}

Now, we can write that a design is optimal if and only if Karush Kuhn Tucker (KKT) optimality conditions hold for problem~$(P_{\vec{\beta}})$.
In fact, we show in Theorem~\ref{theo:DetteMulti} that these KKT conditions are equivalent to a geometric characterization of
$S_{\vec{\beta}}-$optimality, which generalizes the theorem of Dette~\cite{Dette93} to the case of multiresponse experiments.

\begin{proof}[Proof of Theorem~\ref{theo:DetteMulti}]
Since strong duality holds between Problems~$(P_{\vec{\beta}})$ and~$(D_{\vec{\beta}})$, the Karush Kuhn Tucker (KKT) conditions
characterize the optimal variables. We sum up  the KKT conditions here, which stem from the dualization
step of the proof of Theorem~\ref{theo:SoptMulti}: 
\begin{align}
\textrm{(Feasibility)} &\qquad t_k \vec{c_k} = \sum_{i=1}^s A_{(k),i}^T\ \vec{v_{ik}} \label{KKT:feas}\\
&\qquad \sum_{i=1}^s w_i=1 \label{KKT:feas_w}\\
\textrm{(Comp. Slackness)} &\qquad  \mu_i (\Vert \vec{\tilde{\beta}}^{1/2} \odot \vec{v_i} \Vert -w_i)=\vec{0} 
				  \overset{(\textrm{since } \mu_i=1)}{\Longrightarrow} w_i=\Vert \vec{\tilde{\beta}}^{1/2} \odot \vec{v_i} \Vert \label{KKT:CS}\\ 
\textrm{(Stationarity)} &\qquad \beta_k=t_k \vec{h_k}^T \vec{c_k} \label{KKT:stat_ti}\\
%&\qquad \mu_k=1\\ %LAISSER ?
&\qquad\left\lbrace \begin{array}{ll}
         \Vert \vec{\tilde{\beta}}^{-1/2} \odot \vec{z_i}\Vert \leq 1 \textrm{ and } \vec{v_i}=\vec{0} &\; \textrm{if }w_i=0\\
          \Vert \vec{\tilde{\beta}}^{-1/2} \odot \vec{z_i}\Vert = 1 \textrm{ and } \vec{v_i}=w_i\ \vec{\tilde{\beta}}^{-1} \odot \vec{z_i}  &\; 
\textrm{otherwise.}                    \end{array} \right. \label{KKT:stat_vk}
%&\qquad \Vert \diag( \beta^{-1/2}) \vec{z_i}\Vert \leq 1,\\% \quad  \textrm{(with equality whenever }\vec{v_i}\neq0)\\
%&\qquad \textrm{with equality whenever }\vec{v_i}\neq0:\ w_k(\Vert \diag( \beta^{-1/2}) \vec{z_i}\Vert -  1)=0.
%&\qquad \qquad \qquad \qquad \qquad \qquad \Longrightarrow w_k(\Vert \diag( \beta^{-1/2}) \vec{z_i}\Vert -  1)=0.
\end{align}
%Note that the stationarity equations can also be found by making explicit the subgradient (instead of the gradient) of the Lagrangian,
%since the norm function is not differentiable at $0$. This is referred in the literature as \emph{generalized} KKT conditions.
%Ou laisser et mettre une ref ?
Now, let $(\vec{t},V,\vec{w})$ and $H=[\vec{h_1},...,\vec{h_r}]^T$ be a pair of primal and dual solutions
of Problem~$(P_{\vec{\beta}})$--$(D_{\vec{\beta}})$: they
satisfy KKT equations\eqref{KKT:feas}-\eqref{KKT:stat_vk}.
We set $\epsilon_i=\frac{1}{w_i} v_{i}$ whenever $w_i \neq 0$ and $\vec{\epsilon_i}=\vec{0} \in \R^{rl}$ otherwise, 
so that~\eqref{KKT:CS} implies
$$\forall i\in[s],\quad \sum_{k=1}^r \beta_k \Vert \vec{\epsilon_{ik}} \Vert^2=w_i \leq 1$$
and~\eqref{KKT:feas} implies
$$\forall k\in[r],\quad t_k \vec{c_k}=\sum_{i=1}^s  A_{(k),i}^T \vec{v_{ik}}=\sum_{i=1}^s w_i A_{(k),i}^T  \vec{\epsilon_{ik}}.$$
These relations are nothing but conditions $(i)$ and $(ii)$ of Theorem~\eqref{theo:DetteMulti}.
Clearly, the stationarity equation~\eqref{KKT:stat_ti} is the same as condition $(iv)$ of Theorem~\eqref{theo:DetteMulti}.
It remains to show that $(iii)$ holds. Let $D$ be an arbitrary matrix from $\mathcal{D}_{\vec{\beta}}$:
when the regression region is $\mathcal{X}=[s]$, there exists a vector $\vec{\alpha}$ in the unit simplex of $\R^s$
as well as vectors $(\vec{\delta_i}:=[\vec{\delta_{i1}}^T,\ldots,\vec{\delta_{ir}}^T]^T \in \R^{rl})_{i\in [s]}$ satisfying
 $\Vert \vec{\tilde{\beta}}^{1/2} \odot \vec{\delta_i} \Vert \leq 1$
such that $$D=\sum_{i=1}^s \alpha_i \left( \begin{array}{c} \vec{\delta_{i1}}^T A_{(1),i} \\
					\vdots \\
					 \vec{\delta_{ir}}^T A_{(r),i} 
             \end{array} \right).$$
We now prove that $H=[\vec{h_1},...,\vec{h_r}]^T$ is the \emph{direction} of the supporting hyperplane of $\mathcal{D}_{\vec{\beta}}$:
\begin{align*}
\langle D,H \rangle & = \sum_{i,k} \alpha_i  \vec{\delta_{ik}}^T A_{(k),i} \vec{h_k} \\
& = \sum_{i=1}^s \alpha_i \vec{\delta_i}^T \vec{z_i}\\
& = \sum_{i=1}^s \alpha_i (\vec{\tilde{\beta}}^{1/2} \odot \vec{\delta_i})^T  (\vec{\tilde{\beta}}^{-1/2} \odot \vec{z_i})\\
& \leq \sum_{i=1}^s \alpha_i \leq 1,\\
\end{align*}
where the inequality is Cauchy-Schwarz, and we have used the stationarity condition~\eqref{KKT:stat_vk}.
 Finally, $(iii)$ holds since $\operatorname{Diag}(\vec{t}) C$ lies on the boundary of $\mathcal{D_{\vec{\beta}}}$:
$$\langle \operatorname{Diag}(\vec{t})C,H \rangle=\sum_{k=1}^r t_k \vec{c_k}^T \vec{h_k}=\sum_k \beta_k=1.$$
Conversely, assume that conditions $(i)-(iv)$ hold. We set $\vec{v_i}=w_i \vec{\epsilon_i}$, and we show that $(\vec{t},V,\vec{w})$ and $H$
satisfy the KKT equations\eqref{KKT:feas}-\eqref{KKT:stat_vk}. As in the direct
part of this proof, it is straightforward to show that the stationarity equation~\eqref{KKT:stat_ti} holds,
as well as the feasibility condition~\eqref{KKT:feas}.\\
Let us now define the vector $\vec{z_i}$ as in~$(D_{\vec{\beta}})$:
$$\vec{z_i} = \left( \begin{array}{c} A_{(1),i} \vec{h_1} \\
					\vdots \\
					A_{(r),i} \vec{h_r}          
             \end{array} \right).$$
Condition $(iii)$ states that for all vector $\vec{\alpha}$ in the unit simplex of $\R^s$, and for all vectors $(\vec{\delta_i} \in \R^{sl})_{i\in [s]}$ 
satisfying $\Vert \vec{\tilde{\beta}}^{1/2} \odot \vec{\delta_i} \Vert \leq 1$, we have
$$\sum_{ik} \alpha_i \vec{\delta_{ik}}^T A_{(k),i} \vec{h_k} \leq 1.$$
In particular, if $\vec{\alpha}=\vec{e_i}$ is the $i\th$ unit vector of the canonical basis of $\R^s$, and
$\vec{\delta_i}=\frac{\vec{\tilde{\beta}}^{-1} \odot \vec{z_i}}{ \Vert \vec{\tilde{\beta}}^{-1/2} \odot \vec{z_i} \Vert}$, we obtain:
$$\sum_{ik} \alpha_i \vec{\delta_{ik}}^T A_{(k),i} \vec{h_k}=\vec{\delta_i}^T \vec{z_i} 
= \frac{1}{\Vert \vec{\tilde{\beta}}^{-1/2} \odot \vec{z_i} \Vert} (\vec{\tilde{\beta}}^{-1/2} \odot \vec{z_i} )^T(\vec{\tilde{\beta}}^{-1/2} \odot \vec{z_i} )
= \Vert \vec{\tilde{\beta}}^{-1/2} \odot \vec{z_i} \Vert \leq 1,$$
and we have shown the inequality of~\eqref{KKT:stat_vk}.\\
The fact that $\vec{v_i}=0$ when $w_i=0$ is clear from
the way we have defined $\vec{v_i}$, and the complementary slackness equation~\eqref{KKT:CS} also holds in this case.\\
It remains to show that $\vec{w}$ is feasible~\eqref{KKT:feas_w} and that~\eqref{KKT:stat_vk} holds for $w_i>0$.
Note that~\eqref{KKT:stat_vk}
in turn implies the complementary slackness equation~\eqref{KKT:CS}.\par\vspace{12pt}

To this end, we write:
\begin{align*}
1=\sum_{k=1}^r \beta_k=\sum_{k=1}^r t_k \vec{c_k}^T \vec{h_k} &=\sum_{ik} w_i \vec{\epsilon_{ik}}^T A_{(k),i} \vec{h_k}\\
&=\sum_{i=1}^s w_i  \vec{\epsilon_{i}}^T \vec{z_{i}}\\
&=\sum_{i=1}^s w_i (\vec{\tilde{\beta}}^{1/2} \odot \vec{\epsilon_{i}})^T (\vec{\tilde{\beta}}^{-1/2} \odot \vec{z_{i}})\\
&\leq \sum_{i=1}^s w_i \Vert \vec{\tilde{\beta}}^{1/2} \odot \vec{\epsilon_{i}} \Vert\ \Vert \vec{\tilde{\beta}}^{-1/2} \odot \vec{z_{i}} \Vert.
\end{align*}
The latter inequality is Cauchy-Schwarz, and it provides an upper bound which is the (weighted) mean of terms all smaller than $1$. 
We can therefore write 
\begin{equation}
\sum_{i=1}^s w_i \Vert \vec{\tilde{\beta}}^{1/2} \odot \vec{\epsilon_{i}} \Vert\ \Vert \vec{\tilde{\beta}}^{-1/2} \odot  \vec{z_{i}} \Vert=1, \label{weightsumCS}
\end{equation}
and the Cauchy-Schwarz inequality must be an equality whenever $w_i\neq0$, which occurs if and only if
$\vec{\tilde{\beta}}^{1/2} \odot \vec{\epsilon_{i}}$ is proportional to $\vec{\tilde{\beta}}^{-1/2} \odot  \vec{z_i}$. Finally, we must have
$\sum_i w_i=1$, so that $\vec{w}$ is feasible~\eqref{KKT:feas_w}, and each positively weighted term in the
sum~\eqref{weightsumCS} must be $1$:
$$w_i \neq 0 \Longrightarrow \Vert \vec{\tilde{\beta}}^{1/2} \odot \vec{\epsilon_{i}} \Vert\ \Vert \vec{\tilde{\beta}}^{-1/2} \odot  \vec{z_{i}} \Vert=1 \Longrightarrow \left\lbrace \begin{array}{c}
                                                             \Vert  \vec{\tilde{\beta}}^{1/2} \odot \vec{\epsilon_{i}} \Vert =1\\
								\Vert \vec{\tilde{\beta}}^{-1/2} \odot  \vec{z_i} \Vert = 1
                                                                 \end{array} \right. .$$
These two norm constraints further force the coefficient of proportionality between $\vec{\tilde{\beta}}^{1/2} \odot \vec{\epsilon_{i}}$
and $\vec{\tilde{\beta}}^{-1/2} \odot  \vec{z_i}$ to be $1$, so that  $\vec{\epsilon_{i}}=\vec{\tilde{\beta}}^{-1} \odot \vec{z_i}$,
and $\vec{v_i}=w_i \vec{\tilde{\beta}}^{-1} \odot \vec{z_i}$,
which completes the proof.
\end{proof}

\end{document}